\documentclass[aps, twocolumn, prx,longbibliography,superscriptaddress]{revtex4-1}
\usepackage{graphicx}
\usepackage{dcolumn}
\usepackage{natbib}
 \usepackage{xcolor}
 \usepackage[francais]{babel}
\usepackage[caption=false]{subfig}
\usepackage{color}
\usepackage[utf8]{inputenc}
\usepackage[T1]{fontenc}
\usepackage{bm}
\newcommand*{\hham}{\hat{\mathcal{H}}}
\usepackage{amsmath}

\usepackage{amssymb}
\usepackage{amssymb}

\newcommand{\ket}[1]{| {#1} \rangle }
\newcommand{\bra}[1]{\langle {#1} | }

\begin{document}

\title{Spin-dependent Force from an NV center Ensemble on a Microlever}

\author{Maxime Perdriat}
\affiliation{Laboratoire De Physique de l'\'Ecole Normale Sup\'erieure, ENS, PSL, CNRS,
Sorbonne Université, Université de Paris, 24 rue Lhomond, 75005 Paris, France.}

\author{Alrik Durand$^1$}
\author{Louis Chambard$^1$}
\author{Julien Voisin$^1$}

\author{Gabriel H\'etet$^1$}

\begin{abstract}
 We report the observation of spin-dependent force induced by Nitrogen Vacancy (NV) centers embedded in a diamond crystal attached to a tethered oscillator. This result was obtained using a spin-dependent torque generated by a micro-diamond containing billions of NV centers, placed at the end of a commercially available silicon cantilever. 
 Our experiment highlights the intricate interplay between the spin properties of NV centers and the mechanical behavior of the oscillator in the presence of a pure torque, providing a set of tools for exploring quantum effects in state of the art micro-mechanical systems.
\end{abstract}

\maketitle

In recent years, research on optomechanical systems has advanced significantly, with access to quantum regimes enabled by their strong coupling to light in cavities and the development of higher-quality mechanical systems \cite{aspelmeyer2014cavity,barzanjeh2022optomechanics, mercier2021quantum,kotler2021direct}. Coupling of optomechanical platforms to intrinsically nonlinear systems further led to major breakthroughs, notably in the coupling of surface acoustic waves to superconducting qubits \cite{bild2023schrodinger}. The spin of single magnetic defects in solids also have strong potential owing to their long coherence times \cite{rabl2009strong}.  The coupling between a mechanical mode and such defects has been extensively studied within the context of magnetic resonance force microscopy (MRFM), culminating in the mechanical detection of a single spin in 2004 \cite{rugar2004single}. 


Thanks to its ability to be optically polarized at room temperature, combined with its long coherence and lifetime, the Nitrogen Vacancy center (NV) stands out as a natural candidate for spin-mechanics. Numerous NV-based theoretical protocols have been proposed for ground-state cooling \cite{rabl2009strong, rabl2010cooling}, for generating matter-wave interference \cite{rusconi2022spin,yin2013large} and for observing phase transitions in spin ensembles mediated by mechanical motion \cite{ma2017proposal,wei2015magnetic}. 
Various types of mechanical coupling schemes with NV centers have been considered, such as coupling {\it via} magnetic field gradients \cite{rabl2009strong}, magnetic torques \cite{delord2017strong, ma2017proposal} or crystal strain \cite{Ovartchaiyapong, Teissier}. Recent achievements include detection of Brownian and driven mechanical motion of cantilevers using single NV centers \cite{arcizet2011single, kolkowitz2012coherent} and spin-cooling of librational modes of trapped diamonds utilizing an ensemble of NV centers in levitated diamonds \cite{Delord2020, perdriat2022angle}.






In this paper, we report on the spin-dependent force on a tethered oscillator induced by Nitrogen Vacancy (NV) centers embedded in a diamond crystal. This result was obtained by attaching a diamond with a diameter of 40 microns to the end of a silicon cantilever and by driving the flexural mode using the spin dependent torque from the NV centers in a homogeneous magnetic field. 

Our findings provide a new pathway for exploring quantum-to-classical coupling in hybrid systems. In particular, it shows that the fundamental flexural motion of a singly clamped low-Q cantilever can be excited by NV spins without magnetic field gradients. The latter generally bring about magnetic noise \cite{rugar2004single, kolkowitz2012coherent} and to not lend themselves easily to the use of many spins. We also present a method for mitigating spurious microwave noise, providing valuable insights for research on the mechanical detection of the spin of single centers.
Combined with the low temperature capability, such a platform offers prospects for studying  
phase transitions such as proposed in Wei et al. \cite{wei2015magnetic} and to scalably couple many spins through the mechanical degree of freedom.










\section{Experimental set-up}

\subsection{Mechanical mode properties and detection set-up}

\begin{figure*}[t]
    \includegraphics[width=2\columnwidth]{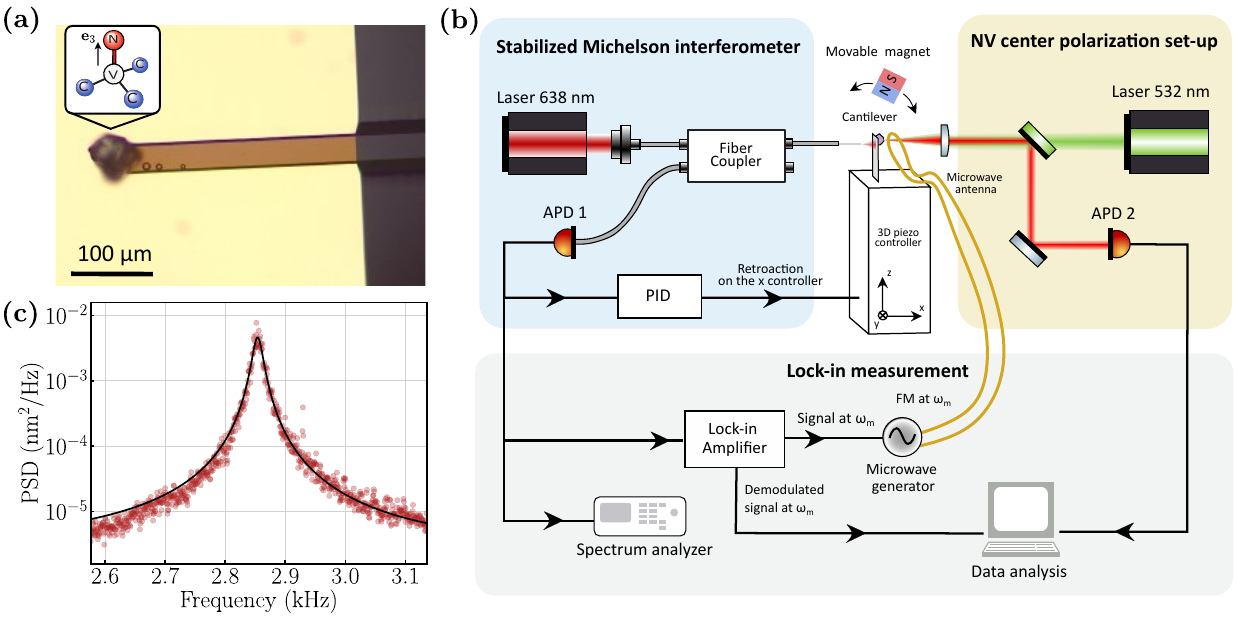}
    \caption{(a) Microscope image of the cantilever with a diamond attached to its extremity. The inset shows the crystallographic structure of the NV center in the diamond.  (b) Schematics of the experiment. A laser at 638 nm (left) is fiber coupled to form an interferometer with the cantilever. The cantilever (in the center) is fixed to a 3-axes controller. A microwave wire is brought close to the diamond attached at the apex of the cantilever. A green laser (right) excites the NV centers inside the diamond and the red photoluminescence is collected through a dichroic mirror and detected on an APD. A 1 inch magnet is used to provide an approximately uniform and tunable magnetic field at the diamond location. (c) Calibrated power spectral density of the fundamental bending mechanical mode of the cantilever.}
    \label{fig:fig_exp_set_up}
\end{figure*}

The mechanical resonator that we consider in this study is the first bending mode of a commercially available Si cantilever from MikroMasch (HQ:CSC38/tipless/No Al). 
A picture of the system is shown in Fig.  \ref{fig:fig_exp_set_up}-(a). 
The dimensions of the cantilever are $350~\mu{\rm m} \times 32.5~\mu{\rm m} \times 1~\mu{\rm m}$. 
We attach a $40~\mu{\rm m}$ diameter diamond bought from Adamas nanotechnology (MDNV40umHi30mg) at the end of the cantilever (see Appendix \ref{app:diamond_positioning} for explanations on the attachment process). 
The cantilever, hereafter referred to as 'loaded,' is enclosed in a vacuum chamber to minimize acoustic noise. However, all experiments presented here were conducted at ambient pressure.

In order to detect the mechanical resonance of the loaded cantilever, we make use of a 
fiber interferometer. The optical setup is shown in Fig. \ref{fig:fig_exp_set_up}-(b).
We use a $638~{\rm nm}$ diode laser that is coupled to an optical fiber and sent to a fiber coupler. One of the outputs of the coupler is directed to a cleaved and stripped fiber positioned in front of the cantilever. The diamond is attached to the other side of the cantilever. The cantilever is fixed to three axes Attocube stages (ANSxyz50 model) in order to position its apex precisely in front of the fiber and to scan it along the optical axis. The light that is reflected off the fiber end (about 15 \% of the incoming power) then interferes with the light that is reflected off the cantilever (about 20 \%). 
The retro-reflected light is detected at the other fiber coupler port by an APD (APD430A from Thorlabs). This electronic signal is then divided into three parts: one part is sent to a PID in order to control the $x$ position of the PZT to remain at half fringe during the measurement. The two other parts of the electric signal are sent to a lock-in amplifier and to a spectrum analyzer.

The resonance frequency of the first bending mode was measured prior to attaching the diamond by recording the power spectral density (PSD) on a spectrum analyzer, with no external excitations other than collisions with the surrounding gas. We found this (unloaded) under-damped frequency mode to be at a frequency $f_0=14.48~{\rm kHz}$, which is consistent with the technical data. The effective mass of the cantilever for this first bending mode can be estimated using the relation $m_{\rm e} \approx \frac{m}{4}$ \cite{rabe1996vibrations}, where $m$ is the total mass of the lever. From the cantilever volume and the density of silicon, we find $m_{\rm e} \approx  10^{-11}~{\rm kg}$.

The calibrated power spectral density of the signal at APD 1 is shown in Fig. \ref{fig:fig_exp_set_up}-(c).
The resonance frequency of this mode is shifted down to $f_{\rm m}=2.86~{\rm kHz}$. 
The mass of a 40~$\mu$m diamond is of the order of $m_{\rm diam}~\approx 10^{-10}~{\rm kg}$, which is ten times larger than the effective mass of the first bending mode. The glue also adds mass to the loaded cantilever. One can estimate the total added mass using  
 the formula $f_{\rm m}=\sqrt{k_{\rm m}/(m_{\rm e}+m_{\rm add})}/2\pi$ for the resonance frequency $f_{\rm m}$ of the first loaded bending mode, where $m_{\rm add}$ is the additional mass of both the glue and the diamond.  We can estimate $m_{\rm add}$ using the formula:
\begin{align}
m_{\rm add}=\left[\left(\frac{f_0}{f_{\rm m}}\right)^2-1 \right] m_{\rm e}.
\end{align}
We find that \( m_{\rm add} \approx 25~ m_{\rm e} \), which is of the right order of magnitude for the mass of the diamond and the glue.

Fitting the PSD using the theoretical mechanical response to background gas collisions in the presence of laser noise gives us a quality factor of $Q=160$ and a stiffness of $k_{\rm m} \approx 0.03$ N/m. To ensure that this measured mode still corresponds to the first bending mode, we perform a mode profile analysis by moving the cantilever along $x$ while recording the thermal displacement for different positions of the cantilever (see Appendix \ref{app:mode_profile}). The measured displacement follows the expected trend of the fundamental flexural mode. 
The thermal displacement is the largest at the apex of the lever and reaches 0.4 nm. This value closely approximates the displacement predicted by the equipartition theorem, assuming the thermal bath is primarily dominated by collisions with a background gas at 300 K.
We can thus estimate the sensitivity of the force detection via the cantilever at this temperature. It reads 
\begin{align}
F_{\rm min}=\sqrt{\frac{4 k_{\rm B} T}{\pi}\frac{k_{\rm m}}{Q f_{\rm m}}}\approx 10^{-14}~{\rm N/\sqrt{\rm Hz}}. 
\label{eq:force_min}
\end{align}
This figure can be improved by employing smaller and textured cantilevers \cite{fischer2019spin} and using lower temperatures. Record sensitivities in the ${\rm zN/\sqrt{\rm Hz}}$ range have for example been achieved with state of the art nanowires at dilution fridge temperatures \cite{Fogliano}.

\subsection{NV centers properties and optical set-up}


 We will now present the properties of the NV centers, which will be used to exert a force on the cantilever.

The unit cell of diamond containing a nitrogen-vacancy (NV) center is illustrated in the inset of Fig. \ref{fig:fig_exp_set_up}-(a). The ground electronic state of the negatively charged NV center is a triplet state, characterized by a zero-field splitting that establishes a natural quantization axis, denoted as $\bm e_3$, along the N-V direction \cite{DOHERTY20131}.  In the absence of an external magnetic field, the $\vert S=1, m_s=\pm 1 \rangle$ states in the triplet manifold have a frequency $D \approx (2\pi) 2.87$ GHz above the $\vert m_s=0 \rangle$ state in the ground state at room temperature. Crucially, the electronic spin of the NV center can be optically polarized to the $\ket{S=1, m_s=0}$ state using green laser light. Additionally, the Stokes-shifted photoluminescence (PL) is stronger in the $\ket{m_s=0}$ state compared to the $\ket{m_s=\pm 1}$ states. Consequently, applying a microwave signal at the resonant frequency induces a drop in PL, which forms the basis of optically detected magnetic resonance (ODMR). In the following, the eigenstates $\ket{m_s=i}$ will simply be denoted as $\ket{i}$. 

The diamond at the end of the cantilever is highly doped with NV centers, with an NV concentration of about $3.5~{\rm ppm}$. Notably, there are four possible orientations of the NV axis in the diamond lattice, as the nitrogen atom can occupy any carbon site adjacent to the vacancy. These four orientations, or classes, yield four distinct projections of the magnetic field along the NV axis. When a magnetic field is applied, the degeneracy between these four classes is lifted, typically revealing eight distinct spin transitions instead of two. Given the NV concentration, we estimate the number of NV centers in a single class to be $N \approx 5\times 10^9$.
\section{Origin of the coupling}

\begin{figure*}[t]
    \includegraphics[width=2.0\columnwidth]{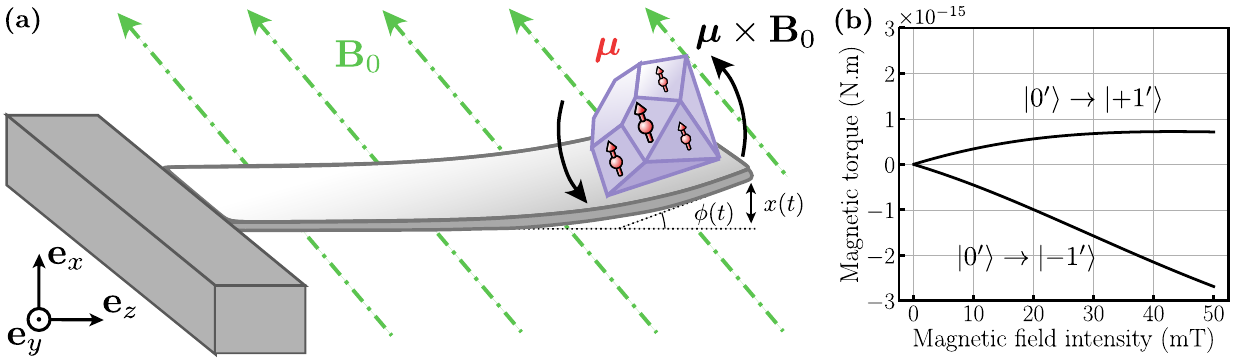}
    \caption{(a) Schematics of the cantilever with a diamond attached to its extremity, depicting how a spin torque in a homogeneous magnetic field enables bending of the cantilever. The torque is here applied along the $\mathbf{e}_y$ direction when a homogeneous magnetic field is applied in the $xz$ plane. (b) Theoretical estimations of the absolute value of the spin torque from the spin of $5 \times 10^9$ NV centers in the diamond. The angle between the magnetic field direction and the NV axis is $60^{\circ}$. }
    \label{fig:fig_cantilever}
\end{figure*}

In this section, we explain the origin of the mechanical coupling between NV centers and the first bending mode of the cantilever in the presence of a homogeneous magnetic field.

\subsection{Cantilever-based torque detection}


In his book ``Foundations of Nanomechanics" \cite{cleland2013foundations}, A.~N. Cleland examines the scenario in which a cantilever is subjected to a constant torque applied perpendicular to its long axis at its tip. It is shown that this not only causes torsion at the end of the cantilever but also induces bending when the finite thickness of the cantilever is taken into account. This treatment demonstrates that bending modes are sensitive to pure torques and can, therefore, be used for torque detection. In recent years, measuring the frequency shift of the first bending mode due to magnetic torques from anisotropic magnetic materials attached to the end of a nanobeam has become a common technique for studying the magnetic properties of these materials \cite{mehlin2015stabilized,gross2016dynamic,gross2020stability,gross2021magnetic}.

By restricting our study in the Fourier domain to dynamical motions that are close to the frequency of the first bending mode $f_{\rm m}$, we can neglect the contribution of other mechanical modes in the displacement of the end of the cantilever $x(t)$. It allows us to write the relation $x(t)=l_{\rm e} \phi(t)$ 
between the angular deflection $\phi(t)$ at the end of the cantilever  and $x(t)$, where $l_{\rm e} \approx 190~\mu {\rm m}$ is the effective length of the first bending mode. The latter is given by the length of the cantilever divided by the mode shape factor $\beta=1.875$. A torque $\tau_{y}$ along $y$ is applied at the end of the cantilever by the NV centers can thus be converted into a force along $\mathbf{e}_x$ via the relation $F_{x}(t)=\tau_{y}(t)/l_{\rm e}$ \cite{gross2016dynamic}. 

Using the above result for our minimal force detection $F_{\rm min}$, we obtain a torque sensitivity
\begin{align}
\tau_{\rm min}\approx 10^{-18}~{\rm N \cdot m/\sqrt{\rm Hz}},
\label{eq:force_min}
\end{align}
which, although orders of magnitude larger than state of the art \cite{Kaviani, Kim}, is enough to measure the magnetic torque applied by the NV centers, as we shall see.


\subsection{Torques induced by NV centers}

The interaction of the negatively charged NV centers in the electronic ground state with magnetic fields can give rise to torques of a different origin.

The spin-spin interaction between the two electrons in the ground state manifold introduces an anisotropy energy term that lifts the degeneracy between the magnetic eigenstates $\ket{\pm 1}$ and the non-magnetic eigenstate $\ket{0}$. This anisotropy term defines a preferential axis for the magnetic moment in the eigenstates $\bm{\mu}_{i}=\mu_{\rm B} \bra{i} \hat{\mathbf{S}} \ket{i}$ to align with. In the presence of an external homogeneous magnetic field $\mathbf{B}_0$, the orientation of $\bm{\mu}_{i}$ results from a competition between both the crystalline anisotropy, and the magnetic field's orientation and strength (see Appendix \ref{app:eigenstates_magnetic_moment} for details). If the magnetic field and the crystalline anisotropy (the N-V axis) are not aligned, then $\bm{\mu}_{i}$ will not be aligned with $\mathbf{B}_0$, resulting in a net magnetic torque $\bm{\mu}_{i} \times \mathbf{B}_0$ applied to the diamond. Finally, the total magnetic torque exerted on the diamond crystal is given by:
\begin{align}
    \bm{\tau}_{\rm mag}=\bm{\mu} \times \mathbf{B}_0,
\end{align}
where $\bm{\mu}=\sum \rho_{ii} \bm{\mu}_{i}$, with $\rho_{ii}$ being the population in the $\ket{i}$ state. This torque has been employed to realize spin-cooling of the libration of levitated diamonds embedded with NV centers \cite{Delord2020}. 
Due to the conservation of angular momentum, any changes in the magnetic moment (proportional to angular momentum) of the NV centers also lead to a change in the macroscopic angular momentum of the diamond crystal. This phenomenon is known as the Einstein-de Haas effect \cite{einstein1915experimental}, and it is responsible for a torque that has notably been observed with yttrium iron garnet (YIG) disks placed on a torsional resonator \cite{mori2020einstein}.
Closely related to this effect, the absorption of microwave photons carrying angular momentum at a magnetic resonance also leads to a net torque on the mechanics. This phenomenon was first observed using a torsional mode of a fiber, where a DPPH crystal (2,2-diphenyl-1-picrylhydrazyl) containing spin one-half paramagnetic impurities was placed at the end \cite{alzetta1967paramagnetic,arimondo1968angular}. For further details on calculations of the NV center torques, see Appendix \ref{app:spin_mecha_calculation}.  
As we will show, only the magnetic torque of the form $\bm{\mu} \times \mathbf{B}_0$ is in line with our experimental observations. 


\begin{figure*}[t]
    \includegraphics[width=2\columnwidth]{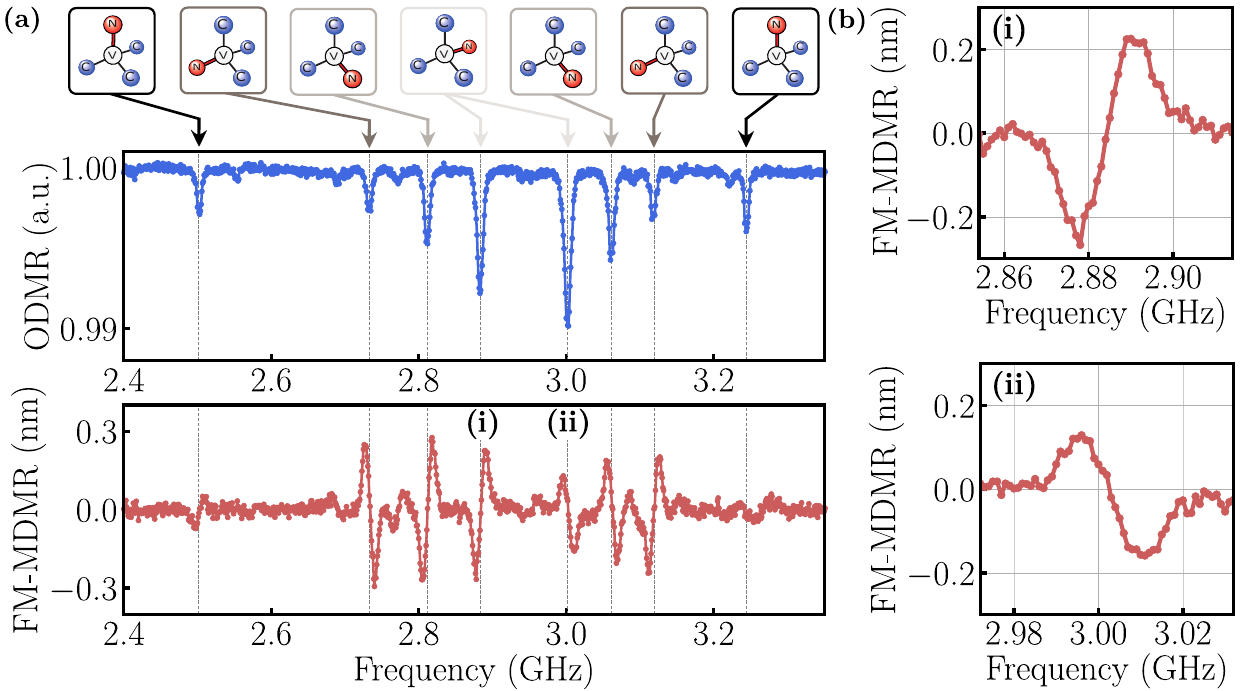}
    \caption{(a) Optically Detected Magnetic Resonance obtained by measuring the photoluminescence from an ensemble of NV centers. The NV orientations are indicated at the top. Bottom curve: Mechanical detection of magnetic resonance using a frequency modulated microwave (FM-MDMR). (b) (i) and (ii): zoom on the FM-MDMR obtained for the same NV orientations for the $\ket{0'} \to \ket{-1'}$ transitions in (i) and for the $\ket{0'} \to \ket{+1'}$ transitions in (ii).}
    \label{fig:fig_exp_magnetic_torque}
\end{figure*}

\subsection{Estimation of the NV centers magnetic torque}

In order to calculate the magnetic torque $\bm{\mu} \times \mathbf{B}_0$ from the NV centers onto the cantilever, we estimate the value of the magnetic moment $\bm{\mu}$ of the NV centers ensemble. Since the NV centers' spins are independent, the total torque is simply the one of a single NV center multiplied by the total number $N$ of NV centers in the diamond. The Hamiltonian of a single NV center reads:
\begin{align}
    \hham_{\rm NV}=\hbar D \left(\hat{\mathbf{S}}\cdot \hat{\mathbf{e}}_3 \right)^2- \hbar \gamma_{\rm e} \mathbf{B}(t) \cdot  \hat{\mathbf{S}},
    \label{eq:hamiltonian_NV}
\end{align}
where $\gamma_{\rm e}=-(2\pi) 28.0~{\rm GHz.T}^{-1}$ is the gyromagnetic factor of the NV center, $D$ is the anisotropy coefficient, $\hat{\mathbf{e}}_3$ is the anisotropy direction of the NV center (see the inset of Fig. \ref{fig:fig_exp_set_up}-(a)), $\mathbf{B}(t)=\mathbf{B}_0+\mathbf{B}_1(t)$ is the sum of an homogeneous magnetic field and a time dependent microwave field. The magnetic field $\mathbf{B}_0$ is chosen not to be aligned with the crystalline anisotropy axis $\hat{\mathbf{e}}_3$, such that a magnetic torque may occur, as previously explained. The microwave field is here employed to drive an electronic spin resonance. We denote $\mathcal{B}=\{\ket{0'},\ket{-1'},\ket{+1'} \}$ as the basis of the new eigenstates of the NV center's Hamiltonian in the magnetic field. Note that the magnetic moment may be non zero in the ground state $\ket{0'}$ because of its mixing with the $\ket{\pm 1}$ magnetic states. This effect can lead to magnetic torques even in the absence of microwaves \cite{pellet2021magnetic, perdriat2023angular}.

Dissipation must also be taken into account to calculate the magnetic moment of the NV centers. The local environment, namely paramagnetic impurities as well as lattice phonons, are responsible for a decay rate $\Gamma_1 \approx (2\pi) 1~{\rm kHz}$ and a dephasing rate $\Gamma_2^*\approx (2\pi) 5~{\rm MHz}$ of the NV centers. Furthermore, the green laser polarizes the NV centers in the $\ket{0'}$ with a rate $\gamma_{\rm las} \approx (2\pi) 1~{\rm kHz}$ in our experiment. This mechanism is efficient if the transverse magnetic field is smaller than 0.05 mT.

By driving a microwave resonant with the magnetic transition $\ket{0'} \to \ket{\pm 1'}$, we can change the magnetic moment, thereby modifying the magnetic torque. The total magnetic moment variation is thus given by $\Delta \bm{\mu}= (\bm{\mu}_{\pm 1'}-\bm{\mu}_{0'})/2$, assuming an initial population in the $\ket{0'}$ state of $100\%$ and a final population of $50-50\%$ in $\ket{0'}$ and $\ket{\pm 1'}$ states.  
Fig. \ref{fig:fig_cantilever}-(b) shows a numerical estimation of the exerted magnetic torque as a function of the magnetic field strength. 
The magnetic field orientation has been chosen such that it makes an angle of $60^{\circ}$ with respect to the NV center axis. The magnetic torque is found to be on the order of $10^{-15}~{\rm N \cdot m}$ at a magnetic field on the order of $10~{\rm mT}$. 
This value is three orders of magnitude above our torque sensitivity.
Notably also, the theoretical torque exerted while driving the $\ket{0'} \to \ket{- 1'}$ transition should be stronger than that of the $\ket{0'}$ to $\ket{+1'}$ transition. 

In the next section, we demonstrate the detection of the motion of the cantilever using the magnetic torque.

\section{Experimental results}

\subsection{Detection of the NV center magnetic torque}

Experimentally, we modulate the torque originating from the NV centers at the mechanical resonance frequency in order to optimise the sensitivity. To achieve this, we perform frequency modulation of the microwave signal at the mechanical frequency while monitoring the interferometer signal, which we hereafter refer to as Frequency Modulated Mechanical Detection of Magnetic Resonance (FM-MDMR).


Fig. \ref{fig:fig_exp_magnetic_torque}-(a) shows an ODMR measurement (blue curve) of the diamond glued to the cantilever. We observe eight dips in the collected photoluminescence, corresponding to the $\ket{0'} \to \ket{-1'}$ and $\ket{0'} \to \ket{+1'}$ transitions of the four NV classes. Other small dips are visible and have been attributed to the presence of an unintended smaller diamond, likely added to the larger mono-crystalline $40~\mu{\rm m}$ diamond during the gluing process. We then scanned the microwave frequency while performing FM-MDMR in the same magnetic field configuration. The modulation amplitude was set to $8~{\rm MHz}$, which is on the order of the NV center linewidth. We measure the in-phase quadrature of the demodulated position signal from the interferometer (this point will be discussed in detail in \ref{sec:sec_dynamics}). 
\begin{figure*}[t]
    \includegraphics[width=2.0\columnwidth]{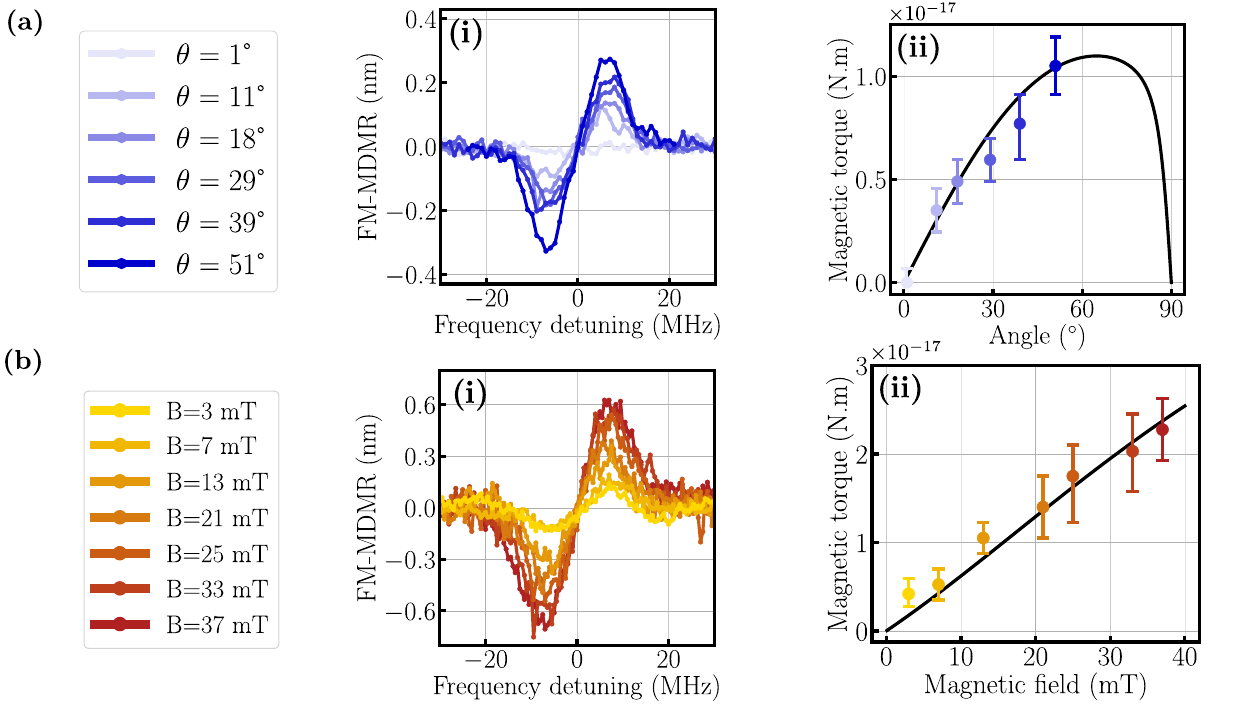}
    \caption{(a) (i) Mechanically-Detected-Magnetic-Resonance with a microwave frequency modulation (FM-MDMR) for different angles between the NV and magnetic field axes at a fixed magnetic field amplitude $B_0=18$ mT. (ii) Magnetic torque as a function of angle extracted from (i). 
    (b) (i) FM-MDMR as a function of magnetic field amplitude. (ii) Magnetic torque as a function of magnetic field amplitude at angle of 70 degrees between the NV and magnetic field axes.  Plain lines are fits to the data by numerical calculations, with the number of polarized spins as a free parameter. Error bars correspond to one standard deviation between the data and the fit.}
    \label{fig:fig_magnetic_field_dependence}
\end{figure*}

The spin-torque measurements are presented in Fig.~\ref{fig:fig_exp_magnetic_torque}-(a) (red curve).
Dispersive signals that all coincide with the ODMR dips are clearly observed, with cantilever displacements on the order of 0.2 nm for the central transitions. 
The dispersive curves are flipped in sign for the $\ket{0'} \to \ket{-1'}$ and $\ket{0'} \to \ket{+1'}$ resonances. This fact is more clearly observed in Fig. \ref{fig:fig_exp_magnetic_torque}-(b) showing a  zoom in on the two resonances of the same class at $2.88~{\rm GHz}$ and $3.00~{\rm GHz}$ respectively. Such opposite behavior indicates that the torque exerted by driving the two transitions are opposite, as expected. Furthermore, the torque exerted by driving the $\ket{0'} \to \ket{+1'}$ transition is consistently smaller than the one on the $\ket{0'} \to \ket{-1'}$ transition. This is in agreement with the theoretical expectations presented in Fig. \ref{fig:fig_cantilever}-(b).  One other possible explanation of the latter observation could be the microwave power dependency on the microwave frequency. The microwave signal power could indeed be lower in the high frequency range corresponding to the $\ket{0'} \to \ket{+1'}$ transitions. 
If it was lower in that range however, the ODMR signals would also show a smaller contrast. 

The torque measured on the cantilever is of the order of $\tau=k_{m} l_{\rm e} x_{\rm res}/Q = 1.0 \times 10^{-17}~{\rm N \cdot m}$, where $x_{\rm res} = 0.2~{\rm nm}$ is the amplitude of the cantilever displacement in FM-MDMR. The measured torque is two orders of magnitude lower than is predicted by the simulation shown in Fig. \ref{fig:fig_cantilever}-(b). This discrepancy may be attributed to multiple factors. First, the green laser polarization  efficiency is not $100\%$, as was assumed in the theoretical calculation. Second, the magnetic torque calculated in Fig. \ref{fig:fig_cantilever}-(b) is the norm of the magnetic torque and not its projection along the $y$-axis. Third, the microwave power was not pushed to its maximum value in order to avoid adding spurious forces given to the cantilever motion. Fourth, we chose the in-phase quadrature of the displacement measurement. The in-phase quadrature is in fact not expected to carry any information about the resonator position 
if the torque were modulated in phase with the mechanical motion. In our case, however, the spin population does not react instantaneously to the frequency modulation, which give some signal to the X quadrature (see \ref{sec:sec_dynamics}). 
Finally, it has been previously observed that diamonds may have a density of NV centers lower than the specification provided by the company. 

\subsection{Dependence of the magnetic torque on the magnetic field}

Although many aspects point towards pure cross-product spin-torques as the origin of the cantilever displacement, it may sill arise from other effects. 
A possible cause of spin dependent cantilever displacement is a magnetic force. The magnetic field generated by the magnet is indeed not perfectly homogeneous, giving rise to magnetic field gradient of the order of $\nabla B \approx 1~{\rm T\cdot m}^{-1}$, leading to a magnetic force of about $10^{-14}~{\rm N}$. 
This force is one order of magnitude lower than the calculated force arising from the magnetic torque, so it is expected to play a minor role in our experiment.
One way to experimentally distinguish a force originating from pure magnetic torque from a force arising due to the magnetic field gradient is by varying the angle $\theta$ between the magnetic field and the NV center axis. A force due to magnetic torque would cancel out when $\theta = 0$, whereas a force caused by the magnetic field gradient would remain unaffected. Fig. \ref{fig:fig_magnetic_field_dependence} (a)-(i) shows FM-MDMR measurements where the angle $\theta$ is varied while keeping the magnetic field at a constant value $B_0=18~{\rm mT}$.
The angles are extracted by diagonalizing the Hamiltonian while varying the magnetic field amplitude and angle to match the energy of the pair of ODMR dips for one NV class. Notably, no mechanical displacement of the cantilever is observed for $\theta \approx  1^\circ$. As observed in Fig. \ref{fig:fig_magnetic_field_dependence} (a)-(ii), the torque indeed increases with $\theta$.


\begin{figure*}[ht]
    \includegraphics[width=2.0\columnwidth]{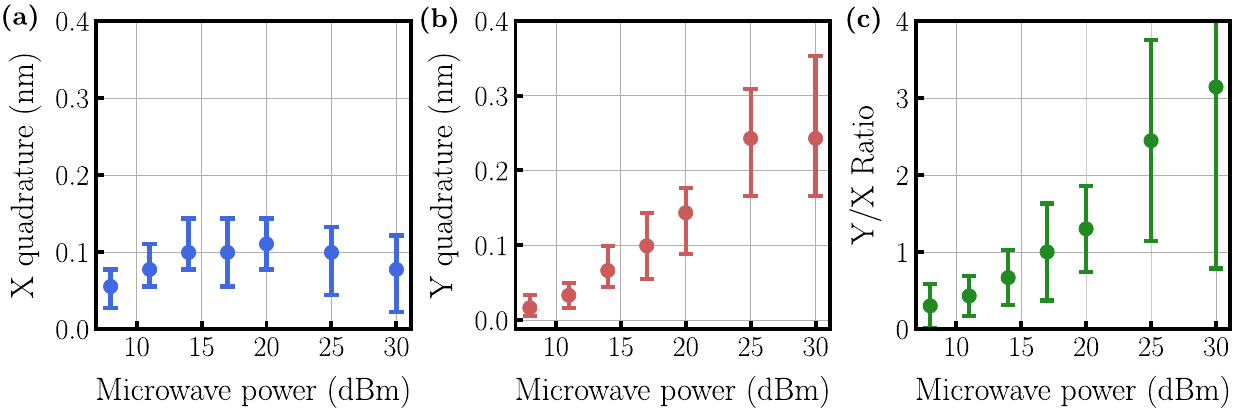}
    \caption{(a) Displacement of the cantilever as a function of microwave power in the X quadrature. (b) Displacement of the cantilever as a function of microwave power in the Y quadrature. (c) Ratio between the two quadratures Y/X as a function of microwave power.}
    \label{fig:fig_X_Y}
\end{figure*}
Another behavior that is also fully consistent with a spin torque coming from a cross-product torque of the form $\bm \mu \times \mathbf{B}_0$ is its dependency with magnetic field amplitude. 
Fig. \ref{fig:fig_magnetic_field_dependence} (b)-(i) shows a FM-MDMR signal obtained for different magnetic field amplitudes, at a fixed angle between the NV axis and the B field. The FM-MDMR signal increases gradually as the B field amplitude is increased. Fig. \ref{fig:fig_magnetic_field_dependence} (b)-(ii) shows the corresponding torque as a function of the magnetic field, featuring close to a linear dependency with magnetic field amplitude. 
\subsection{NV center spin population dynamics}
\label{sec:sec_dynamics}

As previously mentioned, the green laser power ($1~{\rm mW}$) was kept low to avoid affecting the mechanical properties of the cantilever. As a result, the polarization rate is insufficient for the spin population to instantaneously follow the modulation of the microwave frequency. Consequently, the force applied to the mechanics is delayed relative to the frequency modulation excitation by approximately the inverse of the total spin polarization rate, $\gamma_{\rm tot}^{-1}= (\Gamma_1+\gamma_{\rm las} + \frac{\Omega^2}{2\Gamma_2^*})^{-1}$, which is of the order of milliseconds. This value is comparable to the period of the mechanical oscillator considered in our experiment. Therefore, the modulation of the microwave frequency at the mechanical resonance frequency excites both the X and Y quadratures, whereas only the Y quadrature would be expected to respond if the force were not delayed relative to the excitation.

The intensity of the mechanical response in both the X and Y quadratures depends on the polarization rate, and thus on the microwave power. Fig. \ref{fig:fig_X_Y} shows the cantilever displacement as a function of microwave power, measured on the X (a) and Y (b) quadratures of the mechanical oscillator. For the two last points where microwave powers are large, we reduce part of the residual spurious noise by subtracting the FM-MDMR signal obtained with and without green laser polarization (see Appendix \ref{app:soustraction}). It can be observed that the signal in the X quadrature increases slightly, reaching a maximum before decreasing again, while the signal in the Y quadrature steadily increases. The X quadrature response to changes in microwave power can be understood by considering the delay between the spin and the mechanical mode. Up to approximately 15 dBm of applied microwave power, the spin becomes magnetized, causing an increase in torque. However, as the delay decreases, the signal in the X quadrature reaches a maximum and is transferred to the Y quadrature. At 15 dBm, the signals in both quadratures become comparable, although FM-MDMR exhibits several spurious noise peaks, which are absent in the X quadrature. 

Furthermore, the ratio between the Y and X quadrature at the mechanical frequency simply equals 
\begin{align}
Y[\omega_{\rm m}]= \frac{\gamma_{\rm tot}}{\omega_{\rm m}} X[\omega_{\rm m}].
\end{align}
(see Appendix \ref{app:fm_modulation} for the full derivation).
Thus, this ratio is expected to increase linearly with the microwave power. The experimental value of the ratio $Y[\omega_{\rm m}]/X[\omega_{\rm m}]$ as a function of the microwave power plotted in Fig. \ref{fig:fig_X_Y}-(c) t) is consistent with this theoretical prediction.

This delay in the spin population could actually be advantageous in mitigating spurious mechanical motion induced by the microwave signal. 
It has been observed that the microwave signal generates unwanted forces on the cantilever, which are primarily present in the Y quadrature, although the exact origin of these forces is not yet fully understood. 
These forces could be of magnetic, electric, or acoustic origin, potentially stemming from acoustic noise caused by heating of the microwave wire, or from magnetic or electric field gradients generated by the microwave, which exert forces on the slightly diamagnetic/dielectric cantilever. Focusing on the X quadrature, suppresses noise originating from the microwave that is in phase with the microwave frequency modulation.



\section{Conclusion}

In conclusion, we have observed a force induced by the spin of negatively charged NV centers on a clamped micro-mechanical oscillator. This was achieved by frequency modulating the microwave drive at resonance with the cantilever's fundamental flexural mode in the presence of a homogeneous magnetic field. The angular dependence of the force on the direction of the homogeneous magnetic field indicates that the spin torque, via the cross-product, is responsible for coupling to the flexural mode. Homogeneous magnetic fields can be advantageous over gradients, especially when working with spin ensembles larger than a few $\mu$m$^3$, where the inhomogeneous broadening of the spin ensemble becomes excessive due to strong field gradients.





Compared to levitating platforms, one of the advantages of the cantilever approach is that more expensive types of particles can be attached to the cantilever. In particular, at present, the loading efficiency in Paul traps is too low to contemplate using CVD grown micro-diamonds \cite{Perdriat2021}. 
Cantilevers also offer clear prospects for working at larger mechanical frequencies, with foreseeable implications for coherent exchange between spins and mechanics. 

Besides improvements of the sensitivity, future prospects for this work include detecting the Einstein-de Haas effect \cite{mori2020einstein} and probing the dynamics of angular momentum transfer \cite{losby2018recent} at the single spin level.

\begin{acknowledgments}
We thank M. Poggio, O. Arcizet, A. N. Cleland and R. Carpine for enlightening and stimulating discussions. M.P.\ and G.H.\ have been supported by Region Île-de-France in the framework of the DIM Quantip.
This project was funded within the QuantERA II Programme that has received funding from the European Union’s Horizon 2020 research and innovation program under Grant Agreement No 101017733.
\end{acknowledgments}

\appendix

\section{Diamond positioning}
\label{app:diamond_positioning}

To attach the diamond to the cantilever, the process starts by applying a drop of glue to the tip of a metallic wire with a diameter of 15 $\mu$m. Using a 3D micro-metric translation stage, the wire is carefully moved so that the glue touches the end of the cantilever. Next, a single diamond is held by Van der Waals forces at the end of another 15 $\mu$m wire. With the same micrometric stage, the diamond is gently placed onto the glued end of the cantilever. 
All of these steps are performed under an optical microscope. 
The glue is then allowed to dry for a few minutes, securing the diamond in place.

\section{Mode profile of the first bending mode of the cantilever}\label{app:mode_profile}

\begin{figure}
    \includegraphics[width=\columnwidth]{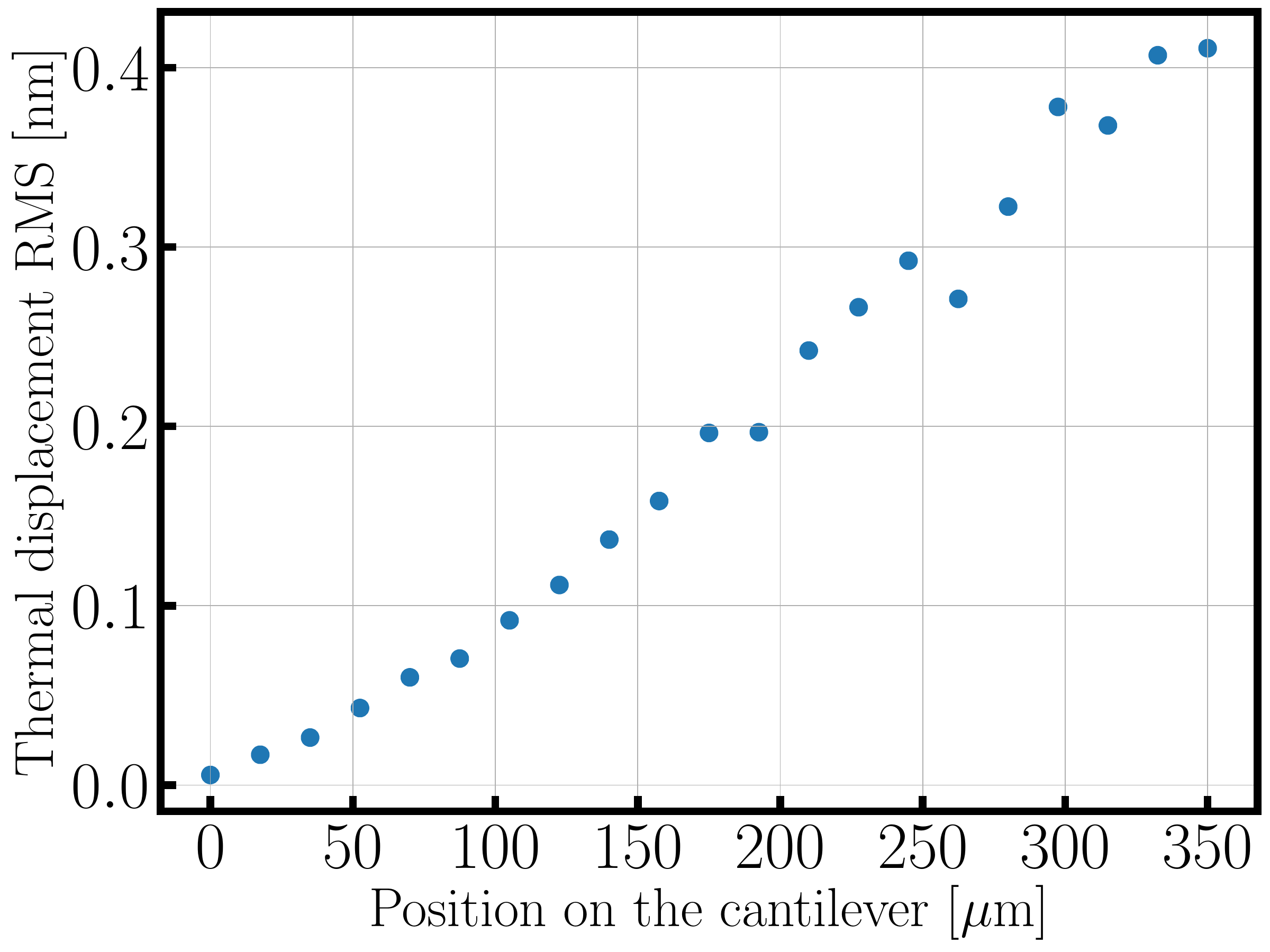}
    \caption{ Thermal displacement of the cantilever
as a function of the laser position on the cantilever. }
    \label{fig:fig_static_result}
\end{figure}

Fig.  \ref{fig:fig_static_result} shows the thermal displacement of the cantilever as a function of the laser position on the cantilever. The displacement reaches 0.4 nm at the cantilever edge and follows a steady trend without nodes, confirming that we indeed measure the first flexural mode.

\section{Estimation of the magnetic moment of the NV center's eigenstates}
\label{app:eigenstates_magnetic_moment}

\begin{figure}[t]
    \includegraphics[width=\columnwidth]{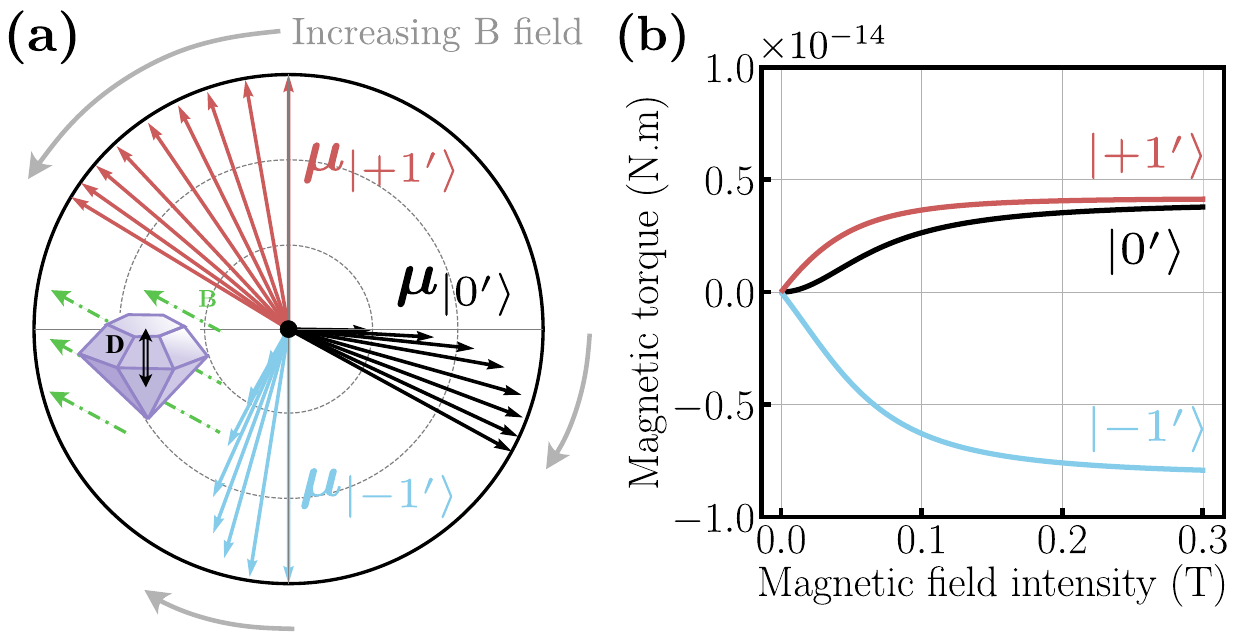}
    \caption{(a) Single spin magnetic moments in the three NV eigenstates as a function of magnetic field with an amplitude ranging from $0.01$ to $0.3$ T, at an angle close to $\pi/4$ with respect to the NV axis as indicated in the figure. (b) Magnetic torque in the three eigenstates as a function of magnetic field strength, ranging from 0 to 0.3 T.}
    \label{fig:fig_magnetic_moment}
\end{figure}

Fig.~\ref{fig:fig_magnetic_moment}(a) presents a calculation of the single-spin magnetic moments of an NV center in its three eigenstates, for different values of the magnetic field strength \( B_0 \). The calculations are performed in a 2D plane defined by \( \mathbf{B}_0 \) and the NV axis \( \mathbf{e}_3 \), assuming that the magnetic field orientation remains fixed at \( 60^\circ \) relative to the anisotropy axis.  

In the low-field regime (\(\gamma_{\rm e} B_0 \ll D\)), the magnetic moment is aligned along the anisotropy axis and remains independent of the magnetic field orientation. In contrast, in the high-field regime (\(\gamma_{\rm e} B_0 \gg D\)), the \(\ket{+1'}\) state aligns with the magnetic field direction, the \(\ket{0'}\) state aligns in the opposite direction, and the \(\ket{-1'}\) state exhibits nearly zero magnetization. In the intermediate regime (\(\gamma_{\rm e} B_0 \approx D\)), the magnetic moment continuously evolves between these two limiting cases.  

The variation in both the orientation and magnitude of the magnetic moment as a function of the magnetic field directly influences the resulting magnetic torque in each of the three eigenstates.
Fig.  \ref{fig:fig_magnetic_moment} (b) shows the result of numerical calculations of the magnetic torque amplitude for a single spin, in each NV eigenstates. We assume here $100\%$ population in each single eigenstate, while keeping the same magnetic field orientation. 
At low magnetic field values, the magnetic torque in the \(\ket{\pm 1'}\) states follows a linear trend, whereas in the \(\ket{0'}\) state, it exhibits a quadratic dependence. For strong magnetic fields, the magnetic torque amplitude in the \(\ket{0'}\) and \(\ket{+1'}\) states converges to a value that is half the opposite of the torque in the \(\ket{-1'}\) state, of the order of \(10^{-14}\)~N·m.  

Interestingly, the magnetic torque remains nonzero even when the magnetization in each eigenstate tends to align with the magnetic field or vanishes. This arises from the fact that the magnetization retains a perpendicular component to the magnetic field, which scales as \(D/B_0\). Consequently, at large magnetic fields, the magnetic torque amplitude is of the order of \(\hbar N D\).  
In practice, the polarization efficiency of the green laser into the \(\ket{0'}\) state becomes ineffective for magnetic field values around \(0.1\)~T.



\section{Calculation of the NV center torques}
\label{app:spin_mecha_calculation}

In this section, we calculate the torques exerted by the NV centers on the diamond in the presence of a homogeneous magnetic field and a microwave field. The Hamiltonian of the spin-mechanical system, involving the rotational degree of freedom of the diamond and an ensemble of NV centers of the same class, is given by:
\begin{align}
\begin{split}
\hham_{\rm s.m.}=\frac{\hat{\mathbf{L}}^2}{2I} +\hbar D \sum_{i=1}^{N} \left(\hat{\mathbf{S}}^{(i)}\cdot \hat{\mathbf{e}}_3\right)^2
- \hbar \gamma_{\rm e} \mathbf{B}(t) \cdot \left(\sum_{i=1}^{N} \hat{\mathbf{S}}^{(i)}\right).
\end{split}
\end{align}
where $\hat{\mathbf{L}}$ is the angular momentum operator of the diamond, $I$ the moment of inertia (the diamond is assumed to be spherical), $D=(2\pi) 2.87~{\rm GHz}$ is the crystalline anisotropy of the NV center, $\hat{\mathbf{S}}^{(i)}$ is the spin-1 operator of the i-th NV center in the diamond, $N$ is the number of NV centers, $\hat{\mathbf{e}}_3$ is the crystalline anisotropy orientation normalized vector, and $\mathbf{B}(t)=\mathbf{B}_0+\mathbf{B}_1(t)$ is the sum of the homogeneous and microwave magnetic field.

Let us now derive the Ehrenfest for the angular momentum mean value $\mathbf{L}=\langle \hat{\mathbf{L}} \rangle$ and the spin operators $\mathbf{S}^{(i)}=\langle \hat{\mathbf{S}}^{(i)} \rangle$:
\begin{align}
\frac{\mathrm{d \mathbf{L}}}{\mathrm{d} t} &= \frac{1}{i\hbar} \left \langle [{\hat{\mathbf{L}}},{\hham_{\rm s.m.}}]\right \rangle,\\
\forall i \in [1,N], \quad \frac{\mathrm{d}\mathbf{S}^{(i)}}{\mathrm{d} t} &= \frac{1}{i\hbar} \left \langle [{\hat{\mathbf{S}}^{(i)}},{\hham_{\rm s.m.}}]\right \rangle.
\end{align}
The angular momentum operator $\hat{\mathbf{L}}$ commutes with $\hat{\mathbf{L}}^2$ and the spin operators $\hat{\mathbf{S}}^{(i)}$. The crystalline anisotropy direction $\hat{\mathbf{e}}_3$ depends on the angular position of the diamond, given by the three Euler angles $(\hat{\alpha},\hat{\beta},\hat{\gamma})$, which do not commute with the angular momentum $\hat{\mathbf{L}}$. This results in a torque exerted by the NV centers on the mechanics. Furthermore, the spin observable $\hat{\mathbf{S}}^{(i)}$ commutes with $\hat{\mathbf{L}}^2$, but it does not commute with the projection of the spin operator onto the vector observable $\hat{\mathbf{e}}_3$ or onto the magnetic field $\mathbf{B}(t)$. These commutators can be calculated (see Appendix B in \cite{perdriat2023angular}) and lead to the following equations:
\begin{align}
\frac{\mathrm{d} \mathbf{L}}{\mathrm{d} t} &= \bm{\tau}_{\rm anis.}, \\
\forall i \in [1,N], \quad \hbar \frac{\mathrm{d}\mathbf{S}^{(i)}}{\mathrm{d} t} &=-\bm{\tau}_{\rm anis.} ^{(i)}+\bm{\tau}_{{\rm mag}}^{(i)},
\end{align}
where $\bm{\tau}_{\rm anis.}^{(i)}$ depends on the anisotropy $D$, the diamond angular position and the spin value $\mathbf{S}^{(i)}$ (see \cite{perdriat2023angular} for the exact expression), $\bm{\tau}_{{\rm mag}}^{(i)}=\hbar \gamma_{\rm e} \mathbf{S}^{(i)} \times \mathbf{B}(t)$ is the magnetic torque, and $\bm{\tau}_{\rm anis.}=\sum_{i=1}^N\bm{\tau}_{\rm anis.} ^{(i)}$. This last relation can be viewed as a consequence of Newton’s third law, the action-reaction principle. Indeed, the torque exerted by the NV center spin ensemble on the diamond crystalline matrix is exactly opposite to the torque exerted by the diamond crystalline matrix on the NV center spins. We now define the total magnetic moment of the NV centers as $\bm{\mu}=\hbar \gamma_{\rm e} \sum_{i=1}^N \mathbf{S}^{(i)}$. We thus obtain the system of equations:
\begin{align}
\frac{\mathrm{d} \mathbf{L}}{\mathrm{d} t} &= \bm{\tau}_{\rm anis.}, \\
\frac{1}{\gamma_{\rm e}}\frac{\mathrm{d}\bm{\mu}}{\mathrm{d} t} &=-\bm{\tau}_{\rm anis.}+\bm{\tau}_{{\rm mag}},
\end{align}
with $\bm{\tau}_{{\rm mag}}=\bm{\mu} \times \mathbf{B}(t)$. Finally, we can obtain the simple dynamical equation:
\begin{align}
\frac{\mathrm{d} \mathbf{L}}{\mathrm{d} t} &= \bm{\mu} \times \mathbf{B}_0 +\bm{\mu} \times \mathbf{B}_1(t) -\frac{1}{\gamma_{\rm e}}\frac{\mathrm{d}\bm{\mu}}{\mathrm{d} t}.
\end{align}
The first torque, $\bm{\mu} \times \mathbf{B}_0$, represents the magnetic torque applied to the diamond due to the homogeneous magnetic field. The second torque $\bm{\mu} \times \mathbf{B}_1(t)$ is the magnetic torque applied to the diamond from the microwave field. The microwave field oscillates at GHz frequencies, so this term could be expected to average out to zero. However, while the magnetic resonance is driven, the spin magnetic moment rotates perpendicular to the microwave field, resulting in a non-zero average magnetic torque. This term is the classical equivalent of the torque resulting from the absorption of circularly polarized microwave photons. Finally, the last torque $-1/\gamma_{\rm e}\mathrm{d}\bm{\mu}/\mathrm{d} t$ corresponds to the Einstein-de Haas torque.

\section{Calculation of the mechanical response to the modulation of the NV center population}\label{app:fm_modulation}
In this section, we determine the mechanical response of the cantilever to the frequency modulation of the microwave. For simplicity, we assume that the \(\ket{0'} \to \ket{\pm 1'}\) resonance is driven on the blue side of the transition. This implies that the microwave detuning is chosen such that  

\[
\Delta(t) = \Gamma_2^* + \delta \omega_\mu(t),
\]

where \(\delta \omega_\mu(t) \ll \Gamma_2^*\) represents the frequency modulation, which is considered small compared to the decoherence rate. While this assumption does not hold in our experiment—since \(\delta \omega_\mu(t)\) is of the order of \(\Gamma_2^*\)—it is made here to simplify the calculations.

\subsection{Dynamical equations of the spin}

We restrict the calculation to the two-level subsystem composed of the $\ket{0'}$ state and either the $\ket{-1'}$ or the $\ket{+1'}$ state. For simplicity of notations, we write $\ket{g}$ the ground state of this two level-system, and $\ket{e}$ the excited state. By considering the density matrix formalism, we can derive the master equation:
\begin{align}
    \frac{\partial \rho_{\rm ee}}{\partial t}&=- \frac{\Gamma_1}{2} (\rho_{\rm ee}-\rho_{\rm gg})-\gamma_{\rm las} \rho_{\rm ee}+i\frac{\Omega}{2}(\rho_{\rm eg}-\rho_{\rm eg}^*),\\
    \frac{\partial \rho_{\rm eg}}{\partial t}&=\left(-\Gamma_2^*+i\Delta(t) \right) \rho_{\rm eg}+i\frac{\Omega}{2}(\rho_{\rm ee}-\rho_{\rm gg}).
\end{align}
where $\rho_{\rm ee}$ (resp. $\rho_{\rm gg}$) is the population in state $\ket{e}$ (resp. $\ket{g}$) verifying $\rho_{\rm ee}+\rho_{\rm gg} \approx 1$, $\rho_{\rm eg}$ are the coherences between these two states in the frame rotating at the energy transitions of the two level-system. $\Omega$ is the Rabi frequency. We neglect the influence of the mechanical displacement in the detuning $\Delta(t)$, since the displacement of the mechanical oscillator is too small to observe back-action from the spin towards the dynamics as was observed in \cite{Delord2020}. 

Using this set of equation, we can estimate the stationary value of the spin population and coherence, as well as the spin dynamical response to the frequency modulation $\delta \omega_\mu(t)$. By injecting the expression $\rho_{\rm ij} (t)= \rho_{\rm ij}^{{(\rm S)}}+\delta \rho_{\rm ij} (t)$ where $\rho_{\rm ij}^{{(\rm S)}}$ is the stationary part and $\delta \rho_{\rm ij} (t)$ the dynamical part, we obtain the value for the stationary values:
\begin{align}
\rho_{\rm ee}^{(\rm S)}&=\frac{1}{2}\left(1-\frac{\gamma_{\rm las}}{\gamma_{\rm tot}}\right),\\
\rho_{\rm gg}^{(\rm S)}&=\frac{1}{2}\left(1+\frac{\gamma_{\rm las}}{\gamma_{\rm tot}}\right),\\
\rho_{\rm eg}^{(\rm S)}&=\frac{-1+i}{2} \frac{\Omega}{2\Gamma_2^*}\left(\rho_{\rm ee}^{(\rm S)}-\rho_{\rm gg}^{(\rm S)}\right),
\end{align}
where $\Gamma_0=\Omega^2/2\Gamma_2^*$ and $\gamma_{\rm tot}=\Gamma_{1}+\gamma_{\rm las}+\Gamma_0$. The dynamical part equations reads:
\begin{align}
    \frac{\partial \delta \rho_{\rm ee}}{\partial t}&=- (\Gamma_1 +\gamma_{\rm las}) \delta \rho_{\rm ee}+i\frac{\Omega}{2}(\delta \rho_{\rm eg}- \delta \rho_{\rm eg}^*),\\
    \frac{\partial \delta \rho_{\rm eg}}{\partial t}&=\left(-\Gamma_2^*+i \Gamma_2^* \right) \delta \rho_{\rm eg}+i \rho_{\rm eg}^{\rm (S)} \delta \omega_\mu+i\Omega \delta \rho_{\rm ee}.
\end{align}
Since $\Gamma_2^*$ is the quickest rate in the equation, we can make the adiabatic approximation and neglect $\frac{\partial \delta \rho_{\rm eg}}{\partial t}$. Thus, we obtain the equation for the population dynamics:
\begin{align}
\frac{\partial \delta \rho_{\rm ee}}{\partial t}=-\gamma_{\rm tot} \delta \rho_{\rm ee}&-\frac{1}{2}\frac{\gamma_{\rm las} \Gamma_0}{\gamma_{\rm tot}}  \frac{ \delta \omega_\mu}{\Gamma_2^*}.
\label{eq:dyn_spin}
\end{align}


\subsection{Dynamic equation of the cantilever mechanical mode}

The mechanical displacement $x(t)$ can be written as a sum of a stationary term and an oscillating term $x(t)=x^{\rm (S)}+\delta x(t)$. We thus obtain the dynamical equation for $\delta x(t)$:
\begin{align}
m_{\rm e} \left( \delta \ddot{x}(t)+\frac{\omega_{\rm m}}{Q} \delta \dot{x}(t)+\omega_{\rm m}^2 \delta x(t) \right)= F_y \delta \rho_{\rm ee} (t).
\label{eq:dyn_mecha}
\end{align}
where $F_y=\frac{\tau_{e,y}-\tau_{g,y}}{l_{\rm e}}$, $\tau_{e,y}$ (resp. $\tau_{g,y}$) is the torque assuming $100\%$ population in the $\ket{e}$ state (resp. $\ket{g}$ state). 

\subsection{Quadratures' amplitudes at the mechanical resonance}

Injecting, \eqref{eq:dyn_spin} into \eqref{eq:dyn_mecha}, we obtain in the Fourier domain:
\begin{eqnarray}\nonumber
\delta x[\omega]&=& - \frac{\gamma_{\rm las}\Gamma_0}{2\gamma_{\rm tot}(\gamma_{\rm tot}+i\omega)}\\ 
&\times& \frac{F_y}{m_{\rm e}\left((\omega_{\rm m}^2-\omega^2)+i\omega\omega_{\rm m}/Q \right)}\frac{\delta \omega_\mu[\omega]}{\Gamma_2^*}. 
\label{eq:dynamic_response}
\end{eqnarray}
The first term in the product in \eqref{eq:dynamic_response} reflects the spin population's response to the frequency modulation, while the second term reflects the mechanical response to an external perturbation. 

We can estimate the in-phase and out-of-phase quadratures $X[\omega]={\rm Re}(\delta x[\omega])$ and $Y[\omega]={\rm Im}(\delta x[\omega])$, evaluated at the mechanical resonance frequency $\omega_{\rm m}$:
\begin{align}
    X[\omega_{\rm m}]&=\frac{F_y Q}{2 m_{\rm e}}\frac{\gamma_{\rm las} \Gamma_0}{\omega_{\rm m}\gamma_{\rm tot}(\gamma_{\rm tot}^2+\omega_{\rm m}^2)}\frac{\delta \omega_\mu[\omega_{\rm m}]}{\Gamma_2^*},\\
    Y[\omega_{\rm m}]&=\frac{F_y Q}{2 m_{\rm e}} \frac{\gamma_{\rm las} \Gamma_0}{\omega_{\rm m}^2(\gamma_{\rm tot}^2+\omega_{\rm m}^2)}\frac{\delta \omega_\mu[\omega_{\rm m}]}{\Gamma_2^*}.
\end{align}
Notably we have the simple relation between the two quadratures:
\begin{align}
Y[\omega_{\rm m}]= \frac{\gamma_{\rm tot}}{\omega_{\rm m}}X[\omega_{\rm m}].
\end{align}
This ratio is an affine function of the microwave power.

\section{FM-MDMR with subtraction of signals obtained with and without green laser}
\label{app:soustraction}

As previously mentioned, undesired effects can introduce spurious noise into the mechanics, especially at high microwave power. In such cases, we can perform clean FM-MDMR measurements by subtracting the signal obtained with the green laser from the signal obtained without it. This technique helps remove noise components that are independent of the green laser.

\begin{figure}[h]
    \includegraphics[width=\columnwidth]{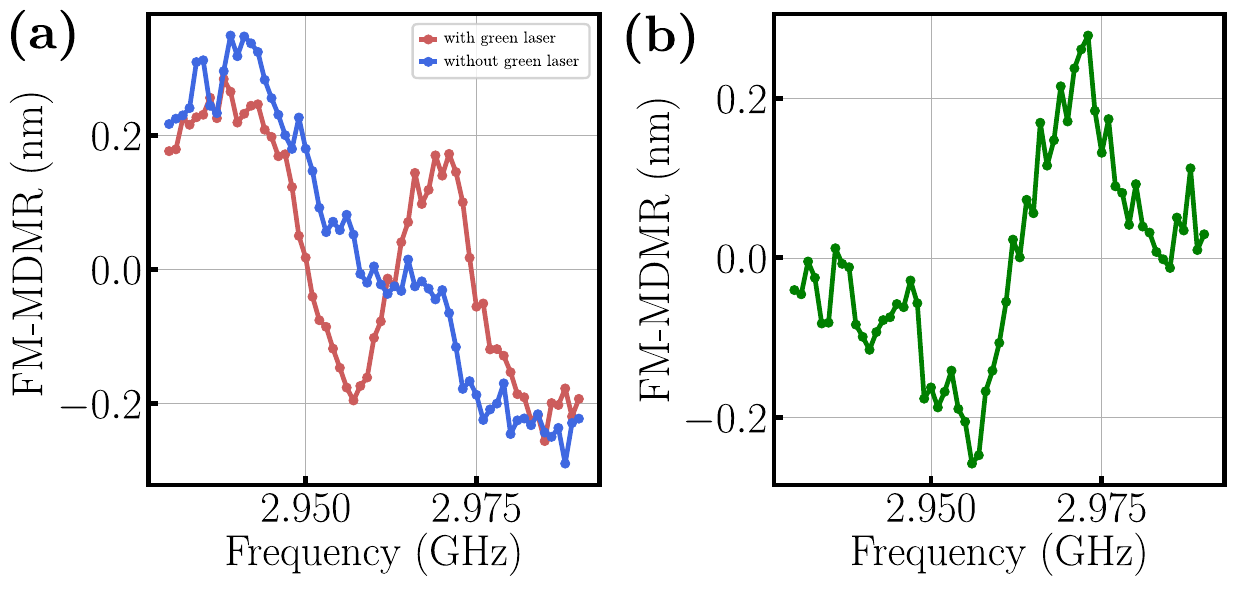}
    \caption{(a) FM-MDMR measured on the $Y$ quadrature as a function of the microwave frequency under green laser excitation (red curve) and without green laser excitation (blue curve). (b) Subtracted FM-MDMR as a function of the microwave frequency.}
    \label{fig:fig_soustraction}
\end{figure}

In Fig.  \ref{fig:fig_soustraction}-(a), we present two FM-MDMR scans taken on the $Y$ quadrature at a microwave power of $30~{\rm dBm}$. The red curve shows the cantilever displacement in the presence of a green laser excitation, while the blue curve was obtained without green laser. The magnetic resonance of the spin transition is equal to $2.962~{\rm GHz}$. The FM-MDMR of Fig. \ref{fig:fig_soustraction}-(b), 
shows the subtraction of the two curves. We observe an expected dispersive curve centered on the magnetic resonance frequency. 

\bibliography{References.bib}

\begin{thebibliography}{38}%
\makeatletter
\providecommand \@ifxundefined [1]{%
 \@ifx{#1\undefined}
}%
\providecommand \@ifnum [1]{%
 \ifnum #1\expandafter \@firstoftwo
 \else \expandafter \@secondoftwo
 \fi
}%
\providecommand \@ifx [1]{%
 \ifx #1\expandafter \@firstoftwo
 \else \expandafter \@secondoftwo
 \fi
}%
\providecommand \natexlab [1]{#1}%
\providecommand \enquote  [1]{``#1''}%
\providecommand \bibnamefont  [1]{#1}%
\providecommand \bibfnamefont [1]{#1}%
\providecommand \citenamefont [1]{#1}%
\providecommand \href@noop [0]{\@secondoftwo}%
\providecommand \href [0]{\begingroup \@sanitize@url \@href}%
\providecommand \@href[1]{\@@startlink{#1}\@@href}%
\providecommand \@@href[1]{\endgroup#1\@@endlink}%
\providecommand \@sanitize@url [0]{\catcode `\\12\catcode `\$12\catcode `\&12\catcode `\#12\catcode `\^12\catcode `\_12\catcode `\%12\relax}%
\providecommand \@@startlink[1]{}%
\providecommand \@@endlink[0]{}%
\providecommand \url  [0]{\begingroup\@sanitize@url \@url }%
\providecommand \@url [1]{\endgroup\@href {#1}{\urlprefix }}%
\providecommand \urlprefix  [0]{URL }%
\providecommand \Eprint [0]{\href }%
\providecommand \doibase [0]{http://dx.doi.org/}%
\providecommand \selectlanguage [0]{\@gobble}%
\providecommand \bibinfo  [0]{\@secondoftwo}%
\providecommand \bibfield  [0]{\@secondoftwo}%
\providecommand \translation [1]{[#1]}%
\providecommand \BibitemOpen [0]{}%
\providecommand \bibitemStop [0]{}%
\providecommand \bibitemNoStop [0]{.\EOS\space}%
\providecommand \EOS [0]{\spacefactor3000\relax}%
\providecommand \BibitemShut  [1]{\csname bibitem#1\endcsname}%
\let\auto@bib@innerbib\@empty
\bibitem [{\citenamefont {Aspelmeyer}\ \emph {et~al.}(2014)\citenamefont {Aspelmeyer}, \citenamefont {Kippenberg},\ and\ \citenamefont {Marquardt}}]{aspelmeyer2014cavity}%
  \BibitemOpen
  \bibfield  {author} {\bibinfo {author} {\bibfnamefont {Markus}\ \bibnamefont {Aspelmeyer}}, \bibinfo {author} {\bibfnamefont {Tobias~J}\ \bibnamefont {Kippenberg}}, \ and\ \bibinfo {author} {\bibfnamefont {Florian}\ \bibnamefont {Marquardt}},\ }\bibfield  {title} {\enquote {\bibinfo {title} {Cavity optomechanics},}\ }\href@noop {} {\bibfield  {journal} {\bibinfo  {journal} {Reviews of Modern Physics}\ }\textbf {\bibinfo {volume} {86}},\ \bibinfo {pages} {1391--1452} (\bibinfo {year} {2014})}\BibitemShut {NoStop}%
\bibitem [{\citenamefont {Barzanjeh}\ \emph {et~al.}(2022)\citenamefont {Barzanjeh}, \citenamefont {Xuereb}, \citenamefont {Gr{\"o}blacher}, \citenamefont {Paternostro}, \citenamefont {Regal},\ and\ \citenamefont {Weig}}]{barzanjeh2022optomechanics}%
  \BibitemOpen
  \bibfield  {author} {\bibinfo {author} {\bibfnamefont {Shabir}\ \bibnamefont {Barzanjeh}}, \bibinfo {author} {\bibfnamefont {Andr{\'e}}\ \bibnamefont {Xuereb}}, \bibinfo {author} {\bibfnamefont {Simon}\ \bibnamefont {Gr{\"o}blacher}}, \bibinfo {author} {\bibfnamefont {Mauro}\ \bibnamefont {Paternostro}}, \bibinfo {author} {\bibfnamefont {Cindy~A}\ \bibnamefont {Regal}}, \ and\ \bibinfo {author} {\bibfnamefont {Eva~M}\ \bibnamefont {Weig}},\ }\bibfield  {title} {\enquote {\bibinfo {title} {Optomechanics for quantum technologies},}\ }\href@noop {} {\bibfield  {journal} {\bibinfo  {journal} {Nature Physics}\ }\textbf {\bibinfo {volume} {18}},\ \bibinfo {pages} {15--24} (\bibinfo {year} {2022})}\BibitemShut {NoStop}%
\bibitem [{\citenamefont {Mercier~de L{\'e}pinay}\ \emph {et~al.}(2021)\citenamefont {Mercier~de L{\'e}pinay}, \citenamefont {Ockeloen-Korppi}, \citenamefont {Woolley},\ and\ \citenamefont {Sillanp{\"a}{\"a}}}]{mercier2021quantum}%
  \BibitemOpen
  \bibfield  {author} {\bibinfo {author} {\bibfnamefont {Laure}\ \bibnamefont {Mercier~de L{\'e}pinay}}, \bibinfo {author} {\bibfnamefont {Caspar~F}\ \bibnamefont {Ockeloen-Korppi}}, \bibinfo {author} {\bibfnamefont {Matthew~J}\ \bibnamefont {Woolley}}, \ and\ \bibinfo {author} {\bibfnamefont {Mika~A}\ \bibnamefont {Sillanp{\"a}{\"a}}},\ }\bibfield  {title} {\enquote {\bibinfo {title} {Quantum mechanics--free subsystem with mechanical oscillators},}\ }\href@noop {} {\bibfield  {journal} {\bibinfo  {journal} {Science}\ }\textbf {\bibinfo {volume} {372}},\ \bibinfo {pages} {625--629} (\bibinfo {year} {2021})}\BibitemShut {NoStop}%
\bibitem [{\citenamefont {Kotler}\ \emph {et~al.}(2021)\citenamefont {Kotler}, \citenamefont {Peterson}, \citenamefont {Shojaee}, \citenamefont {Lecocq}, \citenamefont {Cicak}, \citenamefont {Kwiatkowski}, \citenamefont {Geller}, \citenamefont {Glancy}, \citenamefont {Knill}, \citenamefont {Simmonds} \emph {et~al.}}]{kotler2021direct}%
  \BibitemOpen
  \bibfield  {author} {\bibinfo {author} {\bibfnamefont {Shlomi}\ \bibnamefont {Kotler}}, \bibinfo {author} {\bibfnamefont {Gabriel~A}\ \bibnamefont {Peterson}}, \bibinfo {author} {\bibfnamefont {Ezad}\ \bibnamefont {Shojaee}}, \bibinfo {author} {\bibfnamefont {Florent}\ \bibnamefont {Lecocq}}, \bibinfo {author} {\bibfnamefont {Katarina}\ \bibnamefont {Cicak}}, \bibinfo {author} {\bibfnamefont {Alex}\ \bibnamefont {Kwiatkowski}}, \bibinfo {author} {\bibfnamefont {Shawn}\ \bibnamefont {Geller}}, \bibinfo {author} {\bibfnamefont {Scott}\ \bibnamefont {Glancy}}, \bibinfo {author} {\bibfnamefont {Emanuel}\ \bibnamefont {Knill}}, \bibinfo {author} {\bibfnamefont {Raymond~W}\ \bibnamefont {Simmonds}},  \emph {et~al.},\ }\bibfield  {title} {\enquote {\bibinfo {title} {Direct observation of deterministic macroscopic entanglement},}\ }\href@noop {} {\bibfield  {journal} {\bibinfo  {journal} {Science}\ }\textbf {\bibinfo {volume} {372}},\ \bibinfo {pages} {622--625} (\bibinfo {year} {2021})}\BibitemShut {NoStop}%
\bibitem [{\citenamefont {Bild}\ \emph {et~al.}(2023)\citenamefont {Bild}, \citenamefont {Fadel}, \citenamefont {Yang}, \citenamefont {Von~L{\"u}pke}, \citenamefont {Martin}, \citenamefont {Bruno},\ and\ \citenamefont {Chu}}]{bild2023schrodinger}%
  \BibitemOpen
  \bibfield  {author} {\bibinfo {author} {\bibfnamefont {Marius}\ \bibnamefont {Bild}}, \bibinfo {author} {\bibfnamefont {Matteo}\ \bibnamefont {Fadel}}, \bibinfo {author} {\bibfnamefont {Yu}~\bibnamefont {Yang}}, \bibinfo {author} {\bibfnamefont {Uwe}\ \bibnamefont {Von~L{\"u}pke}}, \bibinfo {author} {\bibfnamefont {Phillip}\ \bibnamefont {Martin}}, \bibinfo {author} {\bibfnamefont {Alessandro}\ \bibnamefont {Bruno}}, \ and\ \bibinfo {author} {\bibfnamefont {Yiwen}\ \bibnamefont {Chu}},\ }\bibfield  {title} {\enquote {\bibinfo {title} {Schr{\"o}dinger cat states of a 16-microgram mechanical oscillator},}\ }\href@noop {} {\bibfield  {journal} {\bibinfo  {journal} {Science}\ }\textbf {\bibinfo {volume} {380}},\ \bibinfo {pages} {274--278} (\bibinfo {year} {2023})}\BibitemShut {NoStop}%
\bibitem [{\citenamefont {Rabl}\ \emph {et~al.}(2009)\citenamefont {Rabl}, \citenamefont {Cappellaro}, \citenamefont {Dutt}, \citenamefont {Jiang}, \citenamefont {Maze},\ and\ \citenamefont {Lukin}}]{rabl2009strong}%
  \BibitemOpen
  \bibfield  {author} {\bibinfo {author} {\bibfnamefont {Peter}\ \bibnamefont {Rabl}}, \bibinfo {author} {\bibfnamefont {P}~\bibnamefont {Cappellaro}}, \bibinfo {author} {\bibfnamefont {MV~Gurudev}\ \bibnamefont {Dutt}}, \bibinfo {author} {\bibfnamefont {Liang}\ \bibnamefont {Jiang}}, \bibinfo {author} {\bibfnamefont {JR}~\bibnamefont {Maze}}, \ and\ \bibinfo {author} {\bibfnamefont {Mikhail~D}\ \bibnamefont {Lukin}},\ }\bibfield  {title} {\enquote {\bibinfo {title} {Strong magnetic coupling between an electronic spin qubit and a mechanical resonator},}\ }\href@noop {} {\bibfield  {journal} {\bibinfo  {journal} {Physical Review B}\ }\textbf {\bibinfo {volume} {79}},\ \bibinfo {pages} {041302} (\bibinfo {year} {2009})}\BibitemShut {NoStop}%
\bibitem [{\citenamefont {Rugar}\ \emph {et~al.}(2004)\citenamefont {Rugar}, \citenamefont {Budakian}, \citenamefont {Mamin},\ and\ \citenamefont {Chui}}]{rugar2004single}%
  \BibitemOpen
  \bibfield  {author} {\bibinfo {author} {\bibfnamefont {Daniel}\ \bibnamefont {Rugar}}, \bibinfo {author} {\bibfnamefont {Raffi}\ \bibnamefont {Budakian}}, \bibinfo {author} {\bibfnamefont {HJ}~\bibnamefont {Mamin}}, \ and\ \bibinfo {author} {\bibfnamefont {BW}~\bibnamefont {Chui}},\ }\bibfield  {title} {\enquote {\bibinfo {title} {Single spin detection by magnetic resonance force microscopy},}\ }\href@noop {} {\bibfield  {journal} {\bibinfo  {journal} {Nature}\ }\textbf {\bibinfo {volume} {430}},\ \bibinfo {pages} {329--332} (\bibinfo {year} {2004})}\BibitemShut {NoStop}%
\bibitem [{\citenamefont {Rabl}(2010)}]{rabl2010cooling}%
  \BibitemOpen
  \bibfield  {author} {\bibinfo {author} {\bibfnamefont {P}~\bibnamefont {Rabl}},\ }\bibfield  {title} {\enquote {\bibinfo {title} {Cooling of mechanical motion with a two-level system: The high-temperature regime},}\ }\href@noop {} {\bibfield  {journal} {\bibinfo  {journal} {Physical Review B—Condensed Matter and Materials Physics}\ }\textbf {\bibinfo {volume} {82}},\ \bibinfo {pages} {165320} (\bibinfo {year} {2010})}\BibitemShut {NoStop}%
\bibitem [{\citenamefont {Rusconi}\ \emph {et~al.}(2022)\citenamefont {Rusconi}, \citenamefont {Perdriat}, \citenamefont {H{\'e}tet}, \citenamefont {Romero-Isart},\ and\ \citenamefont {Stickler}}]{rusconi2022spin}%
  \BibitemOpen
  \bibfield  {author} {\bibinfo {author} {\bibfnamefont {Cosimo~C}\ \bibnamefont {Rusconi}}, \bibinfo {author} {\bibfnamefont {Maxime}\ \bibnamefont {Perdriat}}, \bibinfo {author} {\bibfnamefont {Gabriel}\ \bibnamefont {H{\'e}tet}}, \bibinfo {author} {\bibfnamefont {Oriol}\ \bibnamefont {Romero-Isart}}, \ and\ \bibinfo {author} {\bibfnamefont {Benjamin~A}\ \bibnamefont {Stickler}},\ }\bibfield  {title} {\enquote {\bibinfo {title} {Spin-controlled quantum interference of levitated nanorotors},}\ }\href@noop {} {\bibfield  {journal} {\bibinfo  {journal} {Physical Review Letters}\ }\textbf {\bibinfo {volume} {129}},\ \bibinfo {pages} {093605} (\bibinfo {year} {2022})}\BibitemShut {NoStop}%
\bibitem [{\citenamefont {Yin}\ \emph {et~al.}(2013)\citenamefont {Yin}, \citenamefont {Li}, \citenamefont {Zhang},\ and\ \citenamefont {Duan}}]{yin2013large}%
  \BibitemOpen
  \bibfield  {author} {\bibinfo {author} {\bibfnamefont {Zhang-qi}\ \bibnamefont {Yin}}, \bibinfo {author} {\bibfnamefont {Tongcang}\ \bibnamefont {Li}}, \bibinfo {author} {\bibfnamefont {Xiang}\ \bibnamefont {Zhang}}, \ and\ \bibinfo {author} {\bibfnamefont {LM}~\bibnamefont {Duan}},\ }\bibfield  {title} {\enquote {\bibinfo {title} {Large quantum superpositions of a levitated nanodiamond through spin-optomechanical coupling},}\ }\href@noop {} {\bibfield  {journal} {\bibinfo  {journal} {Physical Review A}\ }\textbf {\bibinfo {volume} {88}},\ \bibinfo {pages} {033614} (\bibinfo {year} {2013})}\BibitemShut {NoStop}%
\bibitem [{\citenamefont {Ma}\ \emph {et~al.}(2017)\citenamefont {Ma}, \citenamefont {Hoang}, \citenamefont {Gong}, \citenamefont {Li},\ and\ \citenamefont {Yin}}]{ma2017proposal}%
  \BibitemOpen
  \bibfield  {author} {\bibinfo {author} {\bibfnamefont {Yue}\ \bibnamefont {Ma}}, \bibinfo {author} {\bibfnamefont {Thai~M}\ \bibnamefont {Hoang}}, \bibinfo {author} {\bibfnamefont {Ming}\ \bibnamefont {Gong}}, \bibinfo {author} {\bibfnamefont {Tongcang}\ \bibnamefont {Li}}, \ and\ \bibinfo {author} {\bibfnamefont {Zhang-qi}\ \bibnamefont {Yin}},\ }\bibfield  {title} {\enquote {\bibinfo {title} {Proposal for quantum many-body simulation and torsional matter-wave interferometry with a levitated nanodiamond},}\ }\href@noop {} {\bibfield  {journal} {\bibinfo  {journal} {Physical Review A}\ }\textbf {\bibinfo {volume} {96}},\ \bibinfo {pages} {023827} (\bibinfo {year} {2017})}\BibitemShut {NoStop}%
\bibitem [{\citenamefont {Wei}\ \emph {et~al.}(2015)\citenamefont {Wei}, \citenamefont {Burk}, \citenamefont {Wrachtrup},\ and\ \citenamefont {Liu}}]{wei2015magnetic}%
  \BibitemOpen
  \bibfield  {author} {\bibinfo {author} {\bibfnamefont {Bo-Bo}\ \bibnamefont {Wei}}, \bibinfo {author} {\bibfnamefont {Christian}\ \bibnamefont {Burk}}, \bibinfo {author} {\bibfnamefont {J{\"o}rg}\ \bibnamefont {Wrachtrup}}, \ and\ \bibinfo {author} {\bibfnamefont {Ren-Bao}\ \bibnamefont {Liu}},\ }\bibfield  {title} {\enquote {\bibinfo {title} {Magnetic ordering of nitrogen-vacancy centers in diamond via resonator-mediated coupling},}\ }\href@noop {} {\bibfield  {journal} {\bibinfo  {journal} {EPJ Quantum Technology}\ }\textbf {\bibinfo {volume} {2}},\ \bibinfo {pages} {1--7} (\bibinfo {year} {2015})}\BibitemShut {NoStop}%
\bibitem [{\citenamefont {Delord}\ \emph {et~al.}(2017)\citenamefont {Delord}, \citenamefont {Nicolas}, \citenamefont {Chassagneux},\ and\ \citenamefont {H{\'e}tet}}]{delord2017strong}%
  \BibitemOpen
  \bibfield  {author} {\bibinfo {author} {\bibfnamefont {T}~\bibnamefont {Delord}}, \bibinfo {author} {\bibfnamefont {L}~\bibnamefont {Nicolas}}, \bibinfo {author} {\bibfnamefont {Y}~\bibnamefont {Chassagneux}}, \ and\ \bibinfo {author} {\bibfnamefont {G}~\bibnamefont {H{\'e}tet}},\ }\bibfield  {title} {\enquote {\bibinfo {title} {Strong coupling between a single nitrogen-vacancy spin and the rotational mode of diamonds levitating in an ion trap},}\ }\href@noop {} {\bibfield  {journal} {\bibinfo  {journal} {Physical Review A}\ }\textbf {\bibinfo {volume} {96}},\ \bibinfo {pages} {063810} (\bibinfo {year} {2017})}\BibitemShut {NoStop}%
\bibitem [{\citenamefont {Ovartchaiyapong}\ \emph {et~al.}(2014)\citenamefont {Ovartchaiyapong}, \citenamefont {Lee}, \citenamefont {Myers},\ and\ \citenamefont {Jayich}}]{Ovartchaiyapong}%
  \BibitemOpen
  \bibfield  {author} {\bibinfo {author} {\bibfnamefont {Preeti}\ \bibnamefont {Ovartchaiyapong}}, \bibinfo {author} {\bibfnamefont {Kenneth~W.}\ \bibnamefont {Lee}}, \bibinfo {author} {\bibfnamefont {Bryan~A.}\ \bibnamefont {Myers}}, \ and\ \bibinfo {author} {\bibfnamefont {Ania C.~Bleszynski}\ \bibnamefont {Jayich}},\ }\bibfield  {title} {\enquote {\bibinfo {title} {Dynamic strain-mediated coupling of a single diamond spin to a mechanical resonator},}\ }\href@noop {} {\bibfield  {journal} {\bibinfo  {journal} {Nature Communications}\ }\textbf {\bibinfo {volume} {5}},\ \bibinfo {pages} {4429} (\bibinfo {year} {2014})}\BibitemShut {NoStop}%
\bibitem [{\citenamefont {Teissier}\ \emph {et~al.}(2014)\citenamefont {Teissier}, \citenamefont {Barfuss}, \citenamefont {Appel}, \citenamefont {Neu},\ and\ \citenamefont {Maletinsky}}]{Teissier}%
  \BibitemOpen
  \bibfield  {author} {\bibinfo {author} {\bibfnamefont {J.}~\bibnamefont {Teissier}}, \bibinfo {author} {\bibfnamefont {A.}~\bibnamefont {Barfuss}}, \bibinfo {author} {\bibfnamefont {P.}~\bibnamefont {Appel}}, \bibinfo {author} {\bibfnamefont {E.}~\bibnamefont {Neu}}, \ and\ \bibinfo {author} {\bibfnamefont {P.}~\bibnamefont {Maletinsky}},\ }\bibfield  {title} {\enquote {\bibinfo {title} {Strain coupling of a nitrogen-vacancy center spin to a diamond mechanical oscillator},}\ }\href {\doibase 10.1103/PhysRevLett.113.020503} {\bibfield  {journal} {\bibinfo  {journal} {Phys. Rev. Lett.}\ }\textbf {\bibinfo {volume} {113}},\ \bibinfo {pages} {020503} (\bibinfo {year} {2014})}\BibitemShut {NoStop}%
\bibitem [{\citenamefont {Arcizet}\ \emph {et~al.}(2011)\citenamefont {Arcizet}, \citenamefont {Jacques}, \citenamefont {Siria}, \citenamefont {Poncharal}, \citenamefont {Vincent},\ and\ \citenamefont {Seidelin}}]{arcizet2011single}%
  \BibitemOpen
  \bibfield  {author} {\bibinfo {author} {\bibfnamefont {Olivier}\ \bibnamefont {Arcizet}}, \bibinfo {author} {\bibfnamefont {Vincent}\ \bibnamefont {Jacques}}, \bibinfo {author} {\bibfnamefont {Alessandro}\ \bibnamefont {Siria}}, \bibinfo {author} {\bibfnamefont {Philippe}\ \bibnamefont {Poncharal}}, \bibinfo {author} {\bibfnamefont {Pascal}\ \bibnamefont {Vincent}}, \ and\ \bibinfo {author} {\bibfnamefont {Signe}\ \bibnamefont {Seidelin}},\ }\bibfield  {title} {\enquote {\bibinfo {title} {A single nitrogen-vacancy defect coupled to a nanomechanical oscillator},}\ }\href@noop {} {\bibfield  {journal} {\bibinfo  {journal} {Nature Physics}\ }\textbf {\bibinfo {volume} {7}},\ \bibinfo {pages} {879--883} (\bibinfo {year} {2011})}\BibitemShut {NoStop}%
\bibitem [{\citenamefont {Kolkowitz}\ \emph {et~al.}(2012)\citenamefont {Kolkowitz}, \citenamefont {Bleszynski~Jayich}, \citenamefont {Unterreithmeier}, \citenamefont {Bennett}, \citenamefont {Rabl}, \citenamefont {Harris},\ and\ \citenamefont {Lukin}}]{kolkowitz2012coherent}%
  \BibitemOpen
  \bibfield  {author} {\bibinfo {author} {\bibfnamefont {Shimon}\ \bibnamefont {Kolkowitz}}, \bibinfo {author} {\bibfnamefont {Ania~C}\ \bibnamefont {Bleszynski~Jayich}}, \bibinfo {author} {\bibfnamefont {Quirin~P}\ \bibnamefont {Unterreithmeier}}, \bibinfo {author} {\bibfnamefont {Steven~D}\ \bibnamefont {Bennett}}, \bibinfo {author} {\bibfnamefont {Peter}\ \bibnamefont {Rabl}}, \bibinfo {author} {\bibfnamefont {JGE}\ \bibnamefont {Harris}}, \ and\ \bibinfo {author} {\bibfnamefont {Mikhail~D}\ \bibnamefont {Lukin}},\ }\bibfield  {title} {\enquote {\bibinfo {title} {Coherent sensing of a mechanical resonator with a single-spin qubit},}\ }\href@noop {} {\bibfield  {journal} {\bibinfo  {journal} {Science}\ }\textbf {\bibinfo {volume} {335}},\ \bibinfo {pages} {1603--1606} (\bibinfo {year} {2012})}\BibitemShut {NoStop}%
\bibitem [{\citenamefont {Delord}\ \emph {et~al.}(2020)\citenamefont {Delord}, \citenamefont {Huillery}, \citenamefont {Nicolas},\ and\ \citenamefont {H{\'e}tet}}]{Delord2020}%
  \BibitemOpen
  \bibfield  {author} {\bibinfo {author} {\bibfnamefont {T.}~\bibnamefont {Delord}}, \bibinfo {author} {\bibfnamefont {P.}~\bibnamefont {Huillery}}, \bibinfo {author} {\bibfnamefont {L.}~\bibnamefont {Nicolas}}, \ and\ \bibinfo {author} {\bibfnamefont {G.}~\bibnamefont {H{\'e}tet}},\ }\bibfield  {title} {\enquote {\bibinfo {title} {Spin-cooling of the motion of a trapped diamond},}\ }\href@noop {} {\bibfield  {journal} {\bibinfo  {journal} {Nature}\ }\textbf {\bibinfo {volume} {580}},\ \bibinfo {pages} {56--59} (\bibinfo {year} {2020})}\BibitemShut {NoStop}%
\bibitem [{\citenamefont {Perdriat}\ \emph {et~al.}(2022)\citenamefont {Perdriat}, \citenamefont {Huillery}, \citenamefont {Pellet-Mary},\ and\ \citenamefont {H{\'e}tet}}]{perdriat2022angle}%
  \BibitemOpen
  \bibfield  {author} {\bibinfo {author} {\bibfnamefont {Maxime}\ \bibnamefont {Perdriat}}, \bibinfo {author} {\bibfnamefont {Paul}\ \bibnamefont {Huillery}}, \bibinfo {author} {\bibfnamefont {Cl{\'e}ment}\ \bibnamefont {Pellet-Mary}}, \ and\ \bibinfo {author} {\bibfnamefont {Gabriel}\ \bibnamefont {H{\'e}tet}},\ }\bibfield  {title} {\enquote {\bibinfo {title} {Angle locking of a levitating diamond using spin diamagnetism},}\ }\href@noop {} {\bibfield  {journal} {\bibinfo  {journal} {Physical Review Letters}\ }\textbf {\bibinfo {volume} {128}},\ \bibinfo {pages} {117203} (\bibinfo {year} {2022})}\BibitemShut {NoStop}%
\bibitem [{\citenamefont {Rabe}\ \emph {et~al.}(1996)\citenamefont {Rabe}, \citenamefont {Janser},\ and\ \citenamefont {Arnold}}]{rabe1996vibrations}%
  \BibitemOpen
  \bibfield  {author} {\bibinfo {author} {\bibfnamefont {U}~\bibnamefont {Rabe}}, \bibinfo {author} {\bibfnamefont {K}~\bibnamefont {Janser}}, \ and\ \bibinfo {author} {\bibfnamefont {Wt}~\bibnamefont {Arnold}},\ }\bibfield  {title} {\enquote {\bibinfo {title} {Vibrations of free and surface-coupled atomic force microscope cantilevers: Theory and experiment},}\ }\href@noop {} {\bibfield  {journal} {\bibinfo  {journal} {Review of scientific instruments}\ }\textbf {\bibinfo {volume} {67}},\ \bibinfo {pages} {3281--3293} (\bibinfo {year} {1996})}\BibitemShut {NoStop}%
\bibitem [{\citenamefont {Fischer}\ \emph {et~al.}(2019)\citenamefont {Fischer}, \citenamefont {McNally}, \citenamefont {Reetz}, \citenamefont {Assumpcao}, \citenamefont {Knief}, \citenamefont {Lin},\ and\ \citenamefont {Regal}}]{fischer2019spin}%
  \BibitemOpen
  \bibfield  {author} {\bibinfo {author} {\bibfnamefont {Ran}\ \bibnamefont {Fischer}}, \bibinfo {author} {\bibfnamefont {Dylan~P}\ \bibnamefont {McNally}}, \bibinfo {author} {\bibfnamefont {Chris}\ \bibnamefont {Reetz}}, \bibinfo {author} {\bibfnamefont {Gabriel~GT}\ \bibnamefont {Assumpcao}}, \bibinfo {author} {\bibfnamefont {T}~\bibnamefont {Knief}}, \bibinfo {author} {\bibfnamefont {Yiheng}\ \bibnamefont {Lin}}, \ and\ \bibinfo {author} {\bibfnamefont {Cindy~A}\ \bibnamefont {Regal}},\ }\bibfield  {title} {\enquote {\bibinfo {title} {Spin detection with a micromechanical trampoline: towards magnetic resonance microscopy harnessing cavity optomechanics},}\ }\href@noop {} {\bibfield  {journal} {\bibinfo  {journal} {New Journal of Physics}\ }\textbf {\bibinfo {volume} {21}},\ \bibinfo {pages} {043049} (\bibinfo {year} {2019})}\BibitemShut {NoStop}%
\bibitem [{\citenamefont {Fogliano}\ \emph {et~al.}(2021)\citenamefont {Fogliano}, \citenamefont {Besga}, \citenamefont {Reigue}, \citenamefont {Mercier~de L{\'e}pinay}, \citenamefont {Heringlake}, \citenamefont {Gouriou}, \citenamefont {Eyraud}, \citenamefont {Wernsdorfer}, \citenamefont {Pigeau},\ and\ \citenamefont {Arcizet}}]{Fogliano}%
  \BibitemOpen
  \bibfield  {author} {\bibinfo {author} {\bibfnamefont {Francesco}\ \bibnamefont {Fogliano}}, \bibinfo {author} {\bibfnamefont {Benjamin}\ \bibnamefont {Besga}}, \bibinfo {author} {\bibfnamefont {Antoine}\ \bibnamefont {Reigue}}, \bibinfo {author} {\bibfnamefont {Laure}\ \bibnamefont {Mercier~de L{\'e}pinay}}, \bibinfo {author} {\bibfnamefont {Philip}\ \bibnamefont {Heringlake}}, \bibinfo {author} {\bibfnamefont {Clement}\ \bibnamefont {Gouriou}}, \bibinfo {author} {\bibfnamefont {Eric}\ \bibnamefont {Eyraud}}, \bibinfo {author} {\bibfnamefont {Wolfgang}\ \bibnamefont {Wernsdorfer}}, \bibinfo {author} {\bibfnamefont {Benjamin}\ \bibnamefont {Pigeau}}, \ and\ \bibinfo {author} {\bibfnamefont {Olivier}\ \bibnamefont {Arcizet}},\ }\bibfield  {title} {\enquote {\bibinfo {title} {Ultrasensitive nano-optomechanical force sensor operated at dilution temperatures},}\ }\href@noop {} {\bibfield  {journal} {\bibinfo  {journal} {Nature Communications}\ }\textbf {\bibinfo {volume} {12}},\ \bibinfo {pages} {4124}
  (\bibinfo {year} {2021})}\BibitemShut {NoStop}%
\bibitem [{\citenamefont {Doherty}\ \emph {et~al.}(2013)\citenamefont {Doherty}, \citenamefont {Manson}, \citenamefont {Delaney}, \citenamefont {Jelezko}, \citenamefont {Wrachtrup},\ and\ \citenamefont {Hollenberg}}]{DOHERTY20131}%
  \BibitemOpen
  \bibfield  {author} {\bibinfo {author} {\bibfnamefont {Marcus~W.}\ \bibnamefont {Doherty}}, \bibinfo {author} {\bibfnamefont {Neil~B.}\ \bibnamefont {Manson}}, \bibinfo {author} {\bibfnamefont {Paul}\ \bibnamefont {Delaney}}, \bibinfo {author} {\bibfnamefont {Fedor}\ \bibnamefont {Jelezko}}, \bibinfo {author} {\bibfnamefont {Jörg}\ \bibnamefont {Wrachtrup}}, \ and\ \bibinfo {author} {\bibfnamefont {Lloyd~C.L.}\ \bibnamefont {Hollenberg}},\ }\bibfield  {title} {\enquote {\bibinfo {title} {The nitrogen-vacancy colour centre in diamond},}\ }\href {\doibase https://doi.org/10.1016/j.physrep.2013.02.001} {\bibfield  {journal} {\bibinfo  {journal} {Physics Reports}\ }\textbf {\bibinfo {volume} {528}},\ \bibinfo {pages} {1--45} (\bibinfo {year} {2013})}\BibitemShut {NoStop}%
\bibitem [{\citenamefont {Cleland}(2013)}]{cleland2013foundations}%
  \BibitemOpen
  \bibfield  {author} {\bibinfo {author} {\bibfnamefont {Andrew~N}\ \bibnamefont {Cleland}},\ }\href@noop {} {\emph {\bibinfo {title} {Foundations of nanomechanics: from solid-state theory to device applications}}}\ (\bibinfo  {publisher} {Springer Science \& Business Media},\ \bibinfo {year} {2013})\BibitemShut {NoStop}%
\bibitem [{\citenamefont {Mehlin}\ \emph {et~al.}(2015)\citenamefont {Mehlin}, \citenamefont {Xue}, \citenamefont {Liang}, \citenamefont {Du}, \citenamefont {Stolt}, \citenamefont {Jin}, \citenamefont {Tian},\ and\ \citenamefont {Poggio}}]{mehlin2015stabilized}%
  \BibitemOpen
  \bibfield  {author} {\bibinfo {author} {\bibfnamefont {A}~\bibnamefont {Mehlin}}, \bibinfo {author} {\bibfnamefont {F}~\bibnamefont {Xue}}, \bibinfo {author} {\bibfnamefont {D}~\bibnamefont {Liang}}, \bibinfo {author} {\bibfnamefont {HF}~\bibnamefont {Du}}, \bibinfo {author} {\bibfnamefont {MJ}~\bibnamefont {Stolt}}, \bibinfo {author} {\bibfnamefont {S}~\bibnamefont {Jin}}, \bibinfo {author} {\bibfnamefont {ML}~\bibnamefont {Tian}}, \ and\ \bibinfo {author} {\bibfnamefont {M}~\bibnamefont {Poggio}},\ }\bibfield  {title} {\enquote {\bibinfo {title} {Stabilized skyrmion phase detected in mnsi nanowires by dynamic cantilever magnetometry},}\ }\href@noop {} {\bibfield  {journal} {\bibinfo  {journal} {Nano letters}\ }\textbf {\bibinfo {volume} {15}},\ \bibinfo {pages} {4839--4844} (\bibinfo {year} {2015})}\BibitemShut {NoStop}%
\bibitem [{\citenamefont {Gross}\ \emph {et~al.}(2016)\citenamefont {Gross}, \citenamefont {Weber}, \citenamefont {R{\"u}ffer}, \citenamefont {Buchter}, \citenamefont {Heimbach}, \citenamefont {Fontcuberta~i Morral}, \citenamefont {Grundler},\ and\ \citenamefont {Poggio}}]{gross2016dynamic}%
  \BibitemOpen
  \bibfield  {author} {\bibinfo {author} {\bibfnamefont {B}~\bibnamefont {Gross}}, \bibinfo {author} {\bibfnamefont {DP}~\bibnamefont {Weber}}, \bibinfo {author} {\bibfnamefont {D}~\bibnamefont {R{\"u}ffer}}, \bibinfo {author} {\bibfnamefont {A}~\bibnamefont {Buchter}}, \bibinfo {author} {\bibfnamefont {F}~\bibnamefont {Heimbach}}, \bibinfo {author} {\bibfnamefont {A}~\bibnamefont {Fontcuberta~i Morral}}, \bibinfo {author} {\bibfnamefont {D}~\bibnamefont {Grundler}}, \ and\ \bibinfo {author} {\bibfnamefont {M}~\bibnamefont {Poggio}},\ }\bibfield  {title} {\enquote {\bibinfo {title} {Dynamic cantilever magnetometry of individual cofeb nanotubes},}\ }\href@noop {} {\bibfield  {journal} {\bibinfo  {journal} {Physical Review B}\ }\textbf {\bibinfo {volume} {93}},\ \bibinfo {pages} {064409} (\bibinfo {year} {2016})}\BibitemShut {NoStop}%
\bibitem [{\citenamefont {Gross}\ \emph {et~al.}(2020)\citenamefont {Gross}, \citenamefont {Philipp}, \citenamefont {Geirhos}, \citenamefont {Mehlin}, \citenamefont {Bord{\'a}cs}, \citenamefont {Tsurkan}, \citenamefont {Leonov}, \citenamefont {K{\'e}zsm{\'a}rki},\ and\ \citenamefont {Poggio}}]{gross2020stability}%
  \BibitemOpen
  \bibfield  {author} {\bibinfo {author} {\bibfnamefont {Boris}\ \bibnamefont {Gross}}, \bibinfo {author} {\bibfnamefont {Simon}\ \bibnamefont {Philipp}}, \bibinfo {author} {\bibfnamefont {Korbinian}\ \bibnamefont {Geirhos}}, \bibinfo {author} {\bibfnamefont {Andrea}\ \bibnamefont {Mehlin}}, \bibinfo {author} {\bibfnamefont {S{\'a}ndor}\ \bibnamefont {Bord{\'a}cs}}, \bibinfo {author} {\bibfnamefont {Vladimir}\ \bibnamefont {Tsurkan}}, \bibinfo {author} {\bibfnamefont {A}~\bibnamefont {Leonov}}, \bibinfo {author} {\bibfnamefont {Istv{\'a}n}\ \bibnamefont {K{\'e}zsm{\'a}rki}}, \ and\ \bibinfo {author} {\bibfnamefont {Martino}\ \bibnamefont {Poggio}},\ }\bibfield  {title} {\enquote {\bibinfo {title} {Stability of n{\'e}el-type skyrmion lattice against oblique magnetic fields in {GaV$_4$S$_8$} and {GaVSe$_8$}},}\ }\href@noop {} {\bibfield  {journal} {\bibinfo  {journal} {Physical Review B}\ }\textbf {\bibinfo {volume} {102}},\ \bibinfo {pages} {104407} (\bibinfo {year} {2020})}\BibitemShut {NoStop}%
\bibitem [{\citenamefont {Gross}\ \emph {et~al.}(2021)\citenamefont {Gross}, \citenamefont {Philipp}, \citenamefont {Josten}, \citenamefont {Leliaert}, \citenamefont {Wetterskog}, \citenamefont {Bergstr{\"o}m},\ and\ \citenamefont {Poggio}}]{gross2021magnetic}%
  \BibitemOpen
  \bibfield  {author} {\bibinfo {author} {\bibfnamefont {B}~\bibnamefont {Gross}}, \bibinfo {author} {\bibfnamefont {S}~\bibnamefont {Philipp}}, \bibinfo {author} {\bibfnamefont {E}~\bibnamefont {Josten}}, \bibinfo {author} {\bibfnamefont {Jonathan}\ \bibnamefont {Leliaert}}, \bibinfo {author} {\bibfnamefont {Erik}\ \bibnamefont {Wetterskog}}, \bibinfo {author} {\bibfnamefont {Lennart}\ \bibnamefont {Bergstr{\"o}m}}, \ and\ \bibinfo {author} {\bibfnamefont {M}~\bibnamefont {Poggio}},\ }\bibfield  {title} {\enquote {\bibinfo {title} {Magnetic anisotropy of individual maghemite mesocrystals},}\ }\href@noop {} {\bibfield  {journal} {\bibinfo  {journal} {Physical Review B}\ }\textbf {\bibinfo {volume} {103}},\ \bibinfo {pages} {014402} (\bibinfo {year} {2021})}\BibitemShut {NoStop}%
\bibitem [{\citenamefont {Kaviani}\ \emph {et~al.}(2023)\citenamefont {Kaviani}, \citenamefont {Behera}, \citenamefont {Hajisalem}, \citenamefont {de~Oliveira~Luiz}, \citenamefont {Lake},\ and\ \citenamefont {Barclay}}]{Kaviani}%
  \BibitemOpen
  \bibfield  {author} {\bibinfo {author} {\bibfnamefont {Hamidreza}\ \bibnamefont {Kaviani}}, \bibinfo {author} {\bibfnamefont {Bishnupada}\ \bibnamefont {Behera}}, \bibinfo {author} {\bibfnamefont {Ghazal}\ \bibnamefont {Hajisalem}}, \bibinfo {author} {\bibfnamefont {Gustavo}\ \bibnamefont {de~Oliveira~Luiz}}, \bibinfo {author} {\bibfnamefont {David~P.}\ \bibnamefont {Lake}}, \ and\ \bibinfo {author} {\bibfnamefont {Paul~E.}\ \bibnamefont {Barclay}},\ }\bibfield  {title} {\enquote {\bibinfo {title} {High-frequency torsional motion transduction using optomechanical coupled oscillators},}\ }\href {\doibase 10.1364/OPTICA.473187} {\bibfield  {journal} {\bibinfo  {journal} {Optica}\ }\textbf {\bibinfo {volume} {10}},\ \bibinfo {pages} {35--41} (\bibinfo {year} {2023})}\BibitemShut {NoStop}%
\bibitem [{\citenamefont {Kim}\ \emph {et~al.}(2016)\citenamefont {Kim}, \citenamefont {Hauer}, \citenamefont {Doolin}, \citenamefont {Souris},\ and\ \citenamefont {Davis}}]{Kim}%
  \BibitemOpen
  \bibfield  {author} {\bibinfo {author} {\bibfnamefont {P.~H.}\ \bibnamefont {Kim}}, \bibinfo {author} {\bibfnamefont {B.~D.}\ \bibnamefont {Hauer}}, \bibinfo {author} {\bibfnamefont {C.}~\bibnamefont {Doolin}}, \bibinfo {author} {\bibfnamefont {F.}~\bibnamefont {Souris}}, \ and\ \bibinfo {author} {\bibfnamefont {J.~P.}\ \bibnamefont {Davis}},\ }\bibfield  {title} {\enquote {\bibinfo {title} {Approaching the standard quantum limit of mechanical torque sensing},}\ }\href@noop {} {\bibfield  {journal} {\bibinfo  {journal} {Nature Communications}\ }\textbf {\bibinfo {volume} {7}},\ \bibinfo {pages} {13165} (\bibinfo {year} {2016})}\BibitemShut {NoStop}%
\bibitem [{\citenamefont {Einstein}\ and\ \citenamefont {De~Haas}(1915)}]{einstein1915experimental}%
  \BibitemOpen
  \bibfield  {author} {\bibinfo {author} {\bibfnamefont {A}~\bibnamefont {Einstein}}\ and\ \bibinfo {author} {\bibfnamefont {WJ}~\bibnamefont {De~Haas}},\ }\bibfield  {title} {\enquote {\bibinfo {title} {Experimental proof of the existence of amp{\`e}re’s molecular currents},}\ }in\ \href@noop {} {\emph {\bibinfo {booktitle} {Proc. KNAW}}},\ Vol.\ \bibinfo {volume} {181}\ (\bibinfo {year} {1915})\ p.\ \bibinfo {pages} {696}\BibitemShut {NoStop}%
\bibitem [{\citenamefont {Mori}\ \emph {et~al.}(2020)\citenamefont {Mori}, \citenamefont {Dunsmore}, \citenamefont {Losby}, \citenamefont {Jenson}, \citenamefont {Belov},\ and\ \citenamefont {Freeman}}]{mori2020einstein}%
  \BibitemOpen
  \bibfield  {author} {\bibinfo {author} {\bibfnamefont {K}~\bibnamefont {Mori}}, \bibinfo {author} {\bibfnamefont {MG}~\bibnamefont {Dunsmore}}, \bibinfo {author} {\bibfnamefont {JE}~\bibnamefont {Losby}}, \bibinfo {author} {\bibfnamefont {DM}~\bibnamefont {Jenson}}, \bibinfo {author} {\bibfnamefont {M}~\bibnamefont {Belov}}, \ and\ \bibinfo {author} {\bibfnamefont {MR}~\bibnamefont {Freeman}},\ }\bibfield  {title} {\enquote {\bibinfo {title} {Einstein--de haas effect at radio frequencies in and near magnetic equilibrium},}\ }\href@noop {} {\bibfield  {journal} {\bibinfo  {journal} {Physical Review B}\ }\textbf {\bibinfo {volume} {102}},\ \bibinfo {pages} {054415} (\bibinfo {year} {2020})}\BibitemShut {NoStop}%
\bibitem [{\citenamefont {Alzetta}\ \emph {et~al.}(1967)\citenamefont {Alzetta}, \citenamefont {Arimondo}, \citenamefont {Ascoli},\ and\ \citenamefont {Gozzini}}]{alzetta1967paramagnetic}%
  \BibitemOpen
  \bibfield  {author} {\bibinfo {author} {\bibfnamefont {G}~\bibnamefont {Alzetta}}, \bibinfo {author} {\bibfnamefont {Ennio}\ \bibnamefont {Arimondo}}, \bibinfo {author} {\bibfnamefont {C}~\bibnamefont {Ascoli}}, \ and\ \bibinfo {author} {\bibfnamefont {A}~\bibnamefont {Gozzini}},\ }\bibfield  {title} {\enquote {\bibinfo {title} {Paramagnetic resonance experiments at low fields with angular-momentum detection},}\ }\href@noop {} {\bibfield  {journal} {\bibinfo  {journal} {Il Nuovo Cimento B (1965-1970)}\ }\textbf {\bibinfo {volume} {52}},\ \bibinfo {pages} {392--402} (\bibinfo {year} {1967})}\BibitemShut {NoStop}%
\bibitem [{\citenamefont {Arimondo}(1968)}]{arimondo1968angular}%
  \BibitemOpen
  \bibfield  {author} {\bibinfo {author} {\bibfnamefont {Ennio}\ \bibnamefont {Arimondo}},\ }\bibfield  {title} {\enquote {\bibinfo {title} {Angular momentum detection of non-linear phenomena in paramagnetic resonance},}\ }in\ \href@noop {} {\emph {\bibinfo {booktitle} {Annales de Physique}}},\ Vol.~\bibinfo {volume} {14}\ (\bibinfo {organization} {EDP Sciences},\ \bibinfo {year} {1968})\ pp.\ \bibinfo {pages} {425--447}\BibitemShut {NoStop}%
\bibitem [{\citenamefont {Pellet-Mary}\ \emph {et~al.}(2021)\citenamefont {Pellet-Mary}, \citenamefont {Huillery}, \citenamefont {Perdriat},\ and\ \citenamefont {H{\'e}tet}}]{pellet2021magnetic}%
  \BibitemOpen
  \bibfield  {author} {\bibinfo {author} {\bibfnamefont {C}~\bibnamefont {Pellet-Mary}}, \bibinfo {author} {\bibfnamefont {Paul}\ \bibnamefont {Huillery}}, \bibinfo {author} {\bibfnamefont {M}~\bibnamefont {Perdriat}}, \ and\ \bibinfo {author} {\bibfnamefont {G}~\bibnamefont {H{\'e}tet}},\ }\bibfield  {title} {\enquote {\bibinfo {title} {Magnetic torque enhanced by tunable dipolar interactions},}\ }\href@noop {} {\bibfield  {journal} {\bibinfo  {journal} {Physical Review B}\ }\textbf {\bibinfo {volume} {104}},\ \bibinfo {pages} {L100411} (\bibinfo {year} {2021})}\BibitemShut {NoStop}%
\bibitem [{\citenamefont {Perdriat}(2023)}]{perdriat2023angular}%
  \BibitemOpen
  \bibfield  {author} {\bibinfo {author} {\bibfnamefont {Maxime}\ \bibnamefont {Perdriat}},\ }\emph {\bibinfo {title} {Angular control of levitated magnetic particles for studies of gyromagnetism at the micro-scale}},\ \href@noop {} {Ph.D. thesis},\ \bibinfo  {school} {Universit{\'e} paris Cit{\'e}} (\bibinfo {year} {2023})\BibitemShut {NoStop}%
\bibitem [{\citenamefont {Perdriat}\ \emph {et~al.}(2021)\citenamefont {Perdriat}, \citenamefont {Pellet-Mary}, \citenamefont {Huillery}, \citenamefont {Rondin},\ and\ \citenamefont {Hétet}}]{Perdriat2021}%
  \BibitemOpen
  \bibfield  {author} {\bibinfo {author} {\bibfnamefont {Maxime}\ \bibnamefont {Perdriat}}, \bibinfo {author} {\bibfnamefont {Clément}\ \bibnamefont {Pellet-Mary}}, \bibinfo {author} {\bibfnamefont {Paul}\ \bibnamefont {Huillery}}, \bibinfo {author} {\bibfnamefont {Loïc}\ \bibnamefont {Rondin}}, \ and\ \bibinfo {author} {\bibfnamefont {Gabriel}\ \bibnamefont {Hétet}},\ }\bibfield  {title} {\enquote {\bibinfo {title} {Spin-mechanics with nitrogen-vacancy centers and trapped particles},}\ }\href {\doibase 10.3390/mi12060651} {\bibfield  {journal} {\bibinfo  {journal} {Micromachines}\ }\textbf {\bibinfo {volume} {12}} (\bibinfo {year} {2021}),\ 10.3390/mi12060651}\BibitemShut {NoStop}%
\bibitem [{\citenamefont {Losby}\ \emph {et~al.}(2018)\citenamefont {Losby}, \citenamefont {Sauer},\ and\ \citenamefont {Freeman}}]{losby2018recent}%
  \BibitemOpen
  \bibfield  {author} {\bibinfo {author} {\bibfnamefont {Joseph~E}\ \bibnamefont {Losby}}, \bibinfo {author} {\bibfnamefont {Vincent~TK}\ \bibnamefont {Sauer}}, \ and\ \bibinfo {author} {\bibfnamefont {Mark~R}\ \bibnamefont {Freeman}},\ }\bibfield  {title} {\enquote {\bibinfo {title} {Recent advances in mechanical torque studies of small-scale magnetism},}\ }\href@noop {} {\bibfield  {journal} {\bibinfo  {journal} {Journal of Physics D: Applied Physics}\ }\textbf {\bibinfo {volume} {51}},\ \bibinfo {pages} {483001} (\bibinfo {year} {2018})}\BibitemShut {NoStop}%
\end{thebibliography}%

\end{document}